\documentclass[
reprint,
superscriptaddress,
amsmath,amssymb,
aps,pra,
longbibliography
]{revtex4-1}

\usepackage{graphicx,xcolor,colortbl}
\newcommand{\snippet}[1]{
    \vspace{10pt}
    \begin{tabular}{!{\color{lightgray}\vrule width 1pt}p{.5\textwidth}}
    \hspace{2pt} \includegraphics[scale=.6]{snippets/#1}
    \end{tabular}
    \vspace{10pt}
}

\usepackage{braket}

\usepackage{xspace} 
\newcommand{\QuESTlink}{QuESTlink\xspace} 
\newcommand{\QuEST}{QuEST\xspace} 

\usepackage[colorlinks = true,
            linkcolor = purple,
            urlcolor  = purple,
            citecolor = purple,
            anchorcolor = purple]{hyperref}

\usepackage{xcolor}
\definecolor{mmablue}{RGB}{1,50,187}
\definecolor{commblue}{rgb}{0.13, 0.67, 0.8}
\definecolor{commgrey}{rgb}{0.6, 0.6, 0.6}
\newcommand{\mmaarg}[1]{\texttt{\textcolor{mmablue}{#1}}}
\newcommand{\mmafunc}[1]{\texttt{#1}}
\newcommand{\mmacomment}[1]{\texttt{\textcolor{commgrey}{#1}}}

\usepackage{tikz}
\usetikzlibrary{shadows}
\newcommand*\keystroke[1]{%
  \tikz[baseline=(key.base)]
    \node[%
      draw,
      fill=white,
      drop shadow={shadow xshift=0.25ex,shadow yshift=-0.25ex,fill=black,opacity=0.75},
      rectangle,
      rounded corners=2pt,
      inner sep=1pt,
      line width=0.5pt,
      font=\scriptsize\sffamily
    ](key) {\;#1\;\strut} ;}
    
\usepackage{enumitem}

\usepackage{titletoc}
    \titlecontents{section}
    [0em] %
    {\medskip\bfseries}
    {\thecontentslabel\quad}
    {}
    {\hfill\contentspage}

    \titlecontents{subsection}[1.72em]{\smallskip}%
    {\thecontentslabel.\enspace}
    {}
    {\titlerule*[1.2pc]{.}\contentspage}%

\begin{document}

\title{QuESTlink -- Mathematica embiggened by a hardware-optimised quantum emulator}

\author{Tyson Jones}
\email{tyson.jones@materials.ox.ac.uk}
\affiliation{Department of Materials, University of Oxford, Parks Road, Oxford OX1 3PH, United Kingdom}
\affiliation{with additional support from Quantum Motion Technologies Ltd UK}
\author{Simon Benjamin}
\email{simon.benjamin@materials.ox.ac.uk}
\affiliation{Department of Materials, University of Oxford, Parks Road, Oxford OX1 3PH, United Kingdom}
\date{\today}


\begin{abstract}
We introduce \href{https://questlink.qtechtheory.org/}{\QuESTlink}, pronounced `quest link', an open-source Mathematica package which efficiently emulates quantum computers. By integrating with the Quantum Exact Simulation Toolkit (\QuEST), \QuESTlink offers a high-level, expressive and usable interface to a high-performance, hardware-accelerated emulator. Requiring no installation, \QuESTlink streamlines the powerful analysis capabilities of Mathematica into the study of quantum systems, even utilising remote multi-core and GPU hardware. We demonstrate the use of \QuESTlink to concisely and efficiently simulate several quantum algorithms, and present some comparative benchmarking against core \QuEST.
\end{abstract}

\maketitle

\tableofcontents


\section{Foreward}

This manuscript uses the phrases ``simulation" and ``emulation" interchangeably, to refer to any use of a classical computer to study, mimic or approximate the behaviour of a digital quantum computer.
While QuESTlink can be classed as a \textit{strong simulator}~\cite{strong_sim},
we use the phrase emulation to distinguish from ``quantum simulation", that is the use of a quantum computer to study a quantum system.


\section{Introduction}

Classical emulation is crucial in the design of quantum computers and algorithms.
Despite the recent demonstration of quantum supremacy~\cite{Google_supremacy}, today's quantum computers are of insufficient quality to run and test many interesting algorithms. Even precise quantum computers of tomorrow may provide limited help in writing new algorithms, since unlike emulators, they offer limited information about the evolving quantum state.
Furthermore, some algorithms, particularly those for noisy intermediate-scale quantum (NISQ) devices~\cite{Preskill2018quantumcomputingin} like the variational class of algorithms~\cite{mcclean2016theory}, admit limited analytic treatment. Hence, the value of classical emulation is undeniable.

The research community needs high-level usable tools that are easy to deploy, offer rapid numerical study, and integrate with other established software.
However, the exponentially-growing cost of classically simulating a quantum device makes emulation of even NISQ computers very resource intensive. Emulators must therefore make good use of classical high-performance computing techniques, like multithreading and GPU parallelisation, and be written in low-level performant languages like C. This requirement is at odds with the need for usable tools, which can be used by non-expert programmers and the wider quantum community.

Within this context we have developed \QuESTlink: a high-performance Mathematica package for numerically emulating quantum computers, by off-loading expensive computation to remote accelerated hardware, running \QuEST~\cite{quest_whitepaper}. Mathematica is both a language and computational tool, prevalent among physicists, which offers a convenient interactive interface (through notebooks), and an extremely comprehensive and powerful set of utilities. Although the most widely used tool for calculations in the physical sciences~\cite{mma_for_physics_book,mma_for_physics_book_2}, Mathematica does not have an intrinsic toolset specifically dedicated to quantum computing emulation. QuESTlink provides a high-performance emulator with a usable Mathematica interface, without compromising the excellent performance and simulation capacity of \QuEST.

From a laptop environment, a user can symbolically specify a circuit in an intuitive high-level operator format akin to how they appear in the literature. Behind a platform-agnostic Mathematica interface, \QuESTlink sends the circuit to a \texttt{C++} backend, either locally or on remote high-performance hardware, where it is efficiently emulated. QuESTlink offers multithreaded and GPU simulation of state vectors and density matrices, multi-qubit multi-controlled general unitaries, general noise processes, circuit drawing, and a wide range of standard gates. Like \QuEST, \QuESTlink is free, open-source and can be used stand-alone; furthermore, without setup, compilation or installation of any kind.


\subsection{\QuESTlink facilities}

While the forthcoming sections give a thorough overview of \QuESTlink's facilities, we here provide a short summary to acquaint the reader with the essentials.
\QuESTlink is, offers, features or enables:

\begin{itemize}[leftmargin=1em]
\setlength\itemsep{0em}
    \item Multithreaded and GPU-accelerated emulation of quantum computers.
    \item State-vector \textit{and} density matrix simulation.
    \item Off-loading of simulation to remote hardware, through a backend-agnostic interface.
    \item Seamless integration with Mathematica's powerful and comprehensive toolset, and interactive notebook programming style.
    \item Rapid development with Mathematica's concise, functional language.
    \item A concise but expressive circuit language, akin to the symbolic description of circuits in the literature.
    \item Stand-alone with no required external downloading, compiling, or installation of any kind whatsoever.
    \item Through Mathematica, is compatible with all major operating systems.
    \item Free and open-source, hosted publicly on Github~\cite{questlink_github}.
    \item Rendering of circuit diagrams, which can be exported to any file format through Mathematica.
    \item A suite of functions for higher-level calculations, like computing Hamiltonian expectation values, and derivatives of unitary circuits.
    \item Analytic generation of matrices from circuit specifications, which can include symbols.
    \item A comprehensive suite of unitary gates and decoherence channels, including multi-controlled multi-qubit general unitaries, and multi-qubit Kraus maps.
    \item A rigorous user-input validation scheme to catch user errors, like supplying non-unitary matrices or physically invalid noise parameters.
\end{itemize}


\section{Technical summary}
\label{sec:technical_summary}

This section provides an overview of the inner workings of \QuESTlink, and the technologies used to build it. Readers intending to use \QuESTlink right away may wish to skip to Section~\ref{sec:how_to_use}.

\subsection{Architecture}

\QuESTlink consists of both a Mathematica package (\mmafunc{QuESTlink.m}) and an underlying \texttt{C++} program (\mmafunc{quest\_link.cpp}), which interface using \texttt{C/Link}; an implementation of the Wolfram Symbolic Transfer Protocol (WSTP)~\cite{WSTP_overview} (formerly called MathLink~\cite{mathlink}). \QuESTlink emulates quantum computers using the Quantum Exact Simulation Toolkit (QuEST)~\cite{quest_whitepaper}, which is an open-source, high-performance emulator written in \texttt{C}. \QuEST is multi-threaded, GPU-accelerated and distributed, and the first two of these facilities are made available to \QuESTlink.

\begin{figure}
    \centering
    \includegraphics[width=.25\textwidth]{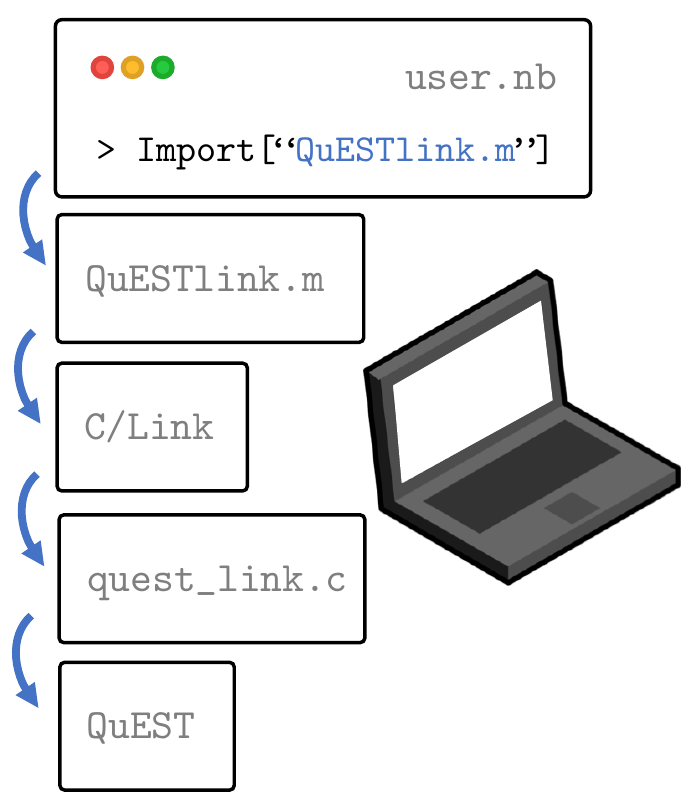}
    \caption{The \QuESTlink software stack, from the user's code (\texttt{user.nb}) to the driving \QuEST~\cite{quest_whitepaper} framework. This stack can run entirely on a user's local machine, achieving an interactive Mathematica interface to a \texttt{C}/\texttt{C++} based emulator. Note the distinction between WSTP (the protocol for communicating between Mathematica and an external process) and \texttt{C/Link} (an implementation for C) is unimportant, and hereafter disregarded.}
    \label{fig:diagram_stack}
\end{figure}

\texttt{C/Link} facilitates conversion of Mathematica types, utilised in the user's notebook, into C types, and maps Mathematica-callable functions to C functions. The executable \texttt{quest\_link} (\texttt{quest\_link.c} compiled with WSTP) sits below as a ``backend" process, performing intermediate processing and invoking \QuEST's API. This software stack is visualised in Figure~\ref{fig:diagram_stack}.

Through this stack, \QuESTlink offers a high-level Mathematica interface to quantum emulation which is numerically performed in \texttt{C}/\texttt{C++} (a significantly faster but lower level language than Mathematica), using memory persistent in the C process, and potentially using accelerated hardware. The expensive numerical representations of the quantum states reside only in the \texttt{C} backend (and through \QuEST, may also be persistent also in GPU memory), and each are identified by a unique ID. Quantum circuits are represented symbolically in their entirety in Mathematica, and sent to the \texttt{C} backend (at a small runtime overhead) only at emulation-time, compactly encoded into arrays of real numbers. The circuit language, and details of the \QuESTlink interface, are presented in Section~\ref{sec:how_to_use}.


QuESTlink leverages QuEST's extensive user-input validation, so that user errors are detected early and reported. For example, this includes validation that input matrices are unitary, that a gate's control and target qubits are unique, and that parameters of a decoherence channel result in a completely positive map. As of QuESTlink v0.3 (integrating QuEST~\href{https://github.com/QuEST-Kit/QuEST/blob/v3.1.0/QuEST/src/QuEST_validation.c}{v3.1.0}), there are 
72 unique forms of input validation. 
This helps users avoid logical errors in their code, to mitigate the risk of drawing incorrect conclusions from the results of QuESTlink.

Core QuEST performs validation before any modification to a quantum register. If a problem is detected, a \texttt{C++} exception is thrown and caught by QuESTlink, and a descriptive error message is propagated to the Mathematica kernel. Some QuESTlink validation is performed \textit{on the fly}. For example, a problem in a gate may only be detected after applying the prior gates in a circuit. In these instances, QuESTlink always reverts the affected register to its original state before the circuit was applied. This means no register is left in an uncertain state.

\subsection{Deployment}

\begin{figure}
    \centering
    \makebox[.1\textwidth][c]{\includegraphics[width=.5\textwidth]{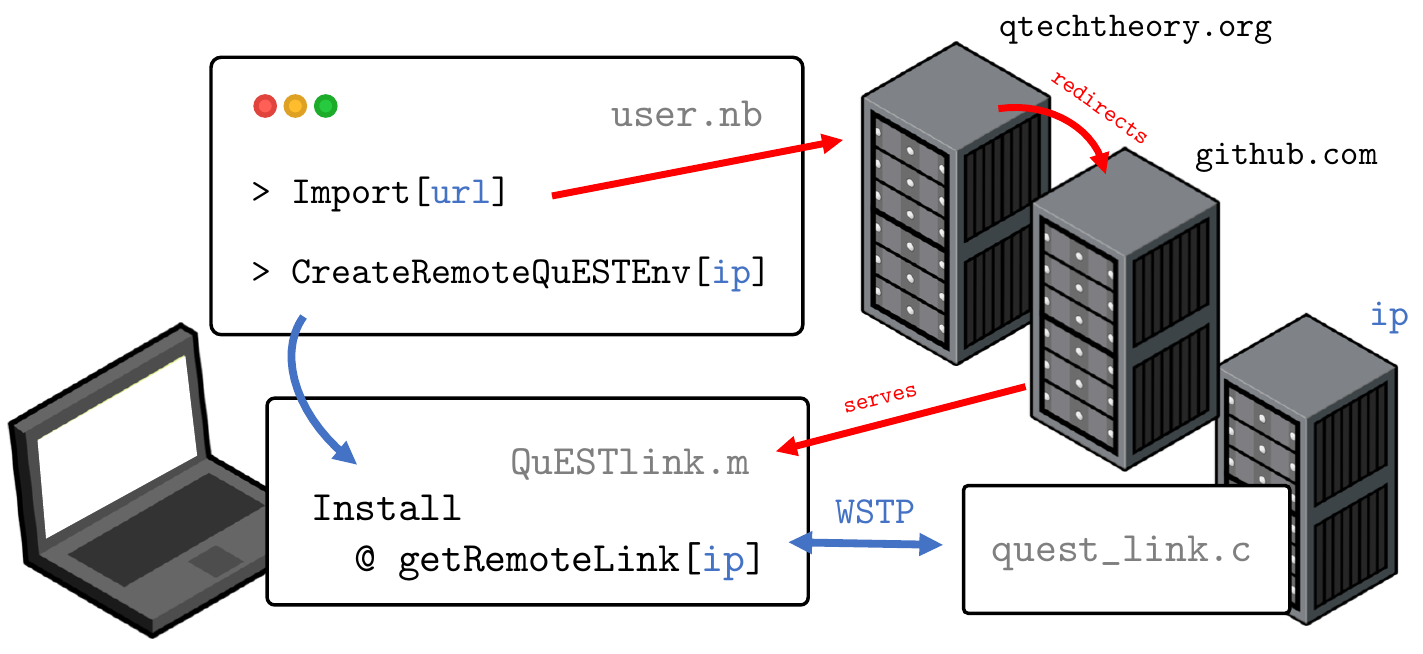}} %
    \caption{The protocol for stand-alone deployment of \QuESTlink.
    First, a user calls Mathematica's \mmafunc{Import[\mmaarg{url}]} function to fetch a copy of the \mmafunc{QuESTlink.m} package, hosted on \href{https://github.com/QTechTheory/QuESTLink}{Github}, but redirected from} \href{https://qtechtheory.org}{qtechtheory.org}.
    This provides the \mmafunc{QuEST'} namespace, and defines functions to connect to a \texttt{quest\_link} backend; this backend can exist on the calling machine, or remotely as here pictured. More details on so called ``\QuEST environments" are presented in detail in the proceeding sections.
    Note the featured code snippets are simplified here for clarity, and their syntax is reviewed in Section~{\ref{sec:mma_syntax_review}.
    }
    \label{fig:diagram_get}
\end{figure}

\QuESTlink can be obtained and deployed entirely within a Mathematica notebook, without any installation or configuration. 
This is done by hosting the package code and \texttt{quest\_link} executable(s) online in a \href{https://github.com/QTechTheory/QuESTlink}{Github}
repository, which the \texttt{qtechtheory.org} server redirects to. Figure~\ref{fig:diagram_get} presents the process of serving the package to a user's Mathematica kernel.

The \texttt{quest\_link} executables for Windows, MacOS and Linux are automatically recompiled on the Github repository with every new release, using Github's integrated continuous-delivery service (\textit{Github Actions}~\cite{github_actions}).
This is for both security, ease of development, and its facility to push bug-fixes and updates seamlessly to the end user.

\subsection{Remote computation}

\begin{figure}
    \centering
    \makebox[.1\textwidth][c]{\includegraphics[width=.5\textwidth]{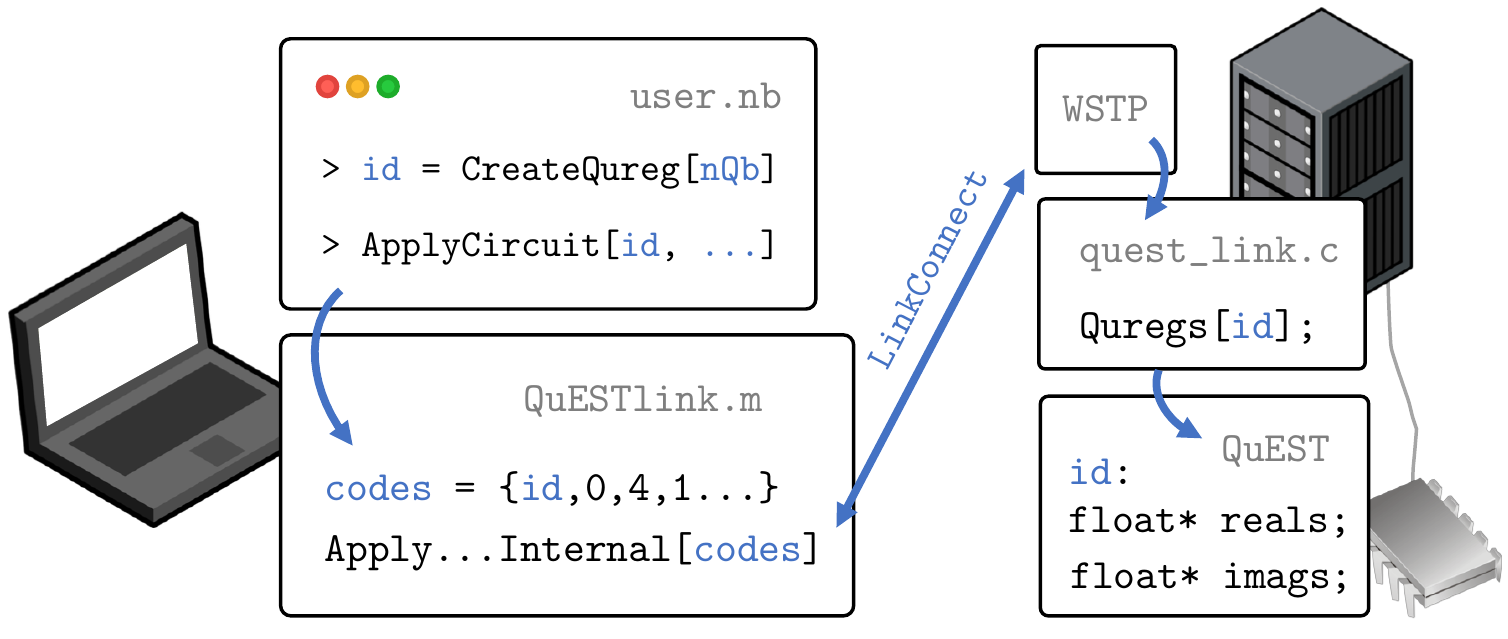}}
    \caption{
    Protocol for off-loading an emulation task onto a remote \QuEST environment. Simulation structures declared in Mathematica are actually allocated remotely, possibly even in accelerated hardware memory.
    Quantum circuits, by first being encoded into arrays of numbers, are communicated to the server over the internet (via TCP/IP).
    Note the syntax used in the code snippets in this diagram are reviewed in Section~{\ref{sec:mma_syntax_review}}
    }
    \label{fig:diagram_run}
\end{figure}

\QuESTlink even enables quantum emulation using \textit{remote} computing resources. Through WSTP, the \texttt{quest\_link} process can run on a remote machine (e.g. a supercomputer) and communicate with the user's local Mathematica kernel via TCP/IP.
The remote machine can employ more powerful hardware than available on the user's machine, and potentially emulate larger quantum states than can fit in the user's local memory. The protocol of off-loading a circuit emulation to remote hardware is outlined in Figure~\ref{fig:diagram_run}. The tools and documentation for setting up a remote \QuESTlink server are provided between the Github repo~\cite{questlink_github}, and \href{https://questlink.qtechtheory.org}{questlink.qtechtheory.org}.

In this remote configuration, Mathematica calls to \QuESTlink functions will involve network communication with the remote environment, and hence incur overheads; this is worthwhile if the remote hardware sufficiently accelerates an expensive emulation which otherwise dominates runtime.

Despite \QuESTlink's effort to minimise the communication cost, this network overhead will be prohibitively costly for some applications. For example, when large simulated quantum states undergo processing by the Mathematica kernel, and hence need to be copied back and forth between the local kernel and the remote environment.
To mitigate this slowdown whilst still running \QuESTlink remotely, one can launch their Mathematica kernel on the remote machine, using Mathematica's remote kernel facilities~\cite{MMA_remote_kernel}. \QuESTlink can also be used inside other Mathematica packages, and so be launched remotely and non-interactively, without a local notebook.


\section{User guide}


\subsection{Mathematica review}
\label{sec:mma_syntax_review}

Before continuing, we offer a quick review of the Mathematica syntax used in this manuscript.

Evaluation of function \mmafunc{f} with input \mmaarg{a} is denoted by
\mmafunc{f[\mmaarg{a}]}, or equivalently \mmafunc{f @ \mmaarg{a}}.
The expression \mmafunc{g[f[\mmaarg{a}]]} can be formed using prefix notation as \mmafunc{g @ f @ \mmaarg{a}}, or postfix notation as \mmafunc{\mmaarg{a} //f //g}.
Matrix multiplication between (possibly complex) matrices \mmaarg{a} and \mmaarg{b} is denoted by \mmafunc{\mmaarg{a}.\mmaarg{b}}, and \mmaarg{a}$^\dagger$ denotes the conjugate transpose of \mmaarg{a}.
Expressions with a trailing \mmafunc{;} suppress their otherwise displayed result. 
While \mmafunc{f[]=\mmaarg{a}} denotes immediate evaluation of \mmaarg{a} and assignment to \mmafunc{f[]}, the syntax \mmafunc{f[]:=\mmaarg{a}} denotes \textit{delayed} assignment, whereby later invoking \mmafunc{f[]} will evaluate \mmaarg{a} (which may have changed) each time.
Elements of a list $\mmafunc{x = }\{\mmaarg{a},\mmaarg{b},\mmaarg{c}\}$ are accessed as \mmafunc{x[[\mmaarg{i}]]} where index $\mmaarg{i} \ge 1$, and \mmafunc{x[[\mmaarg{m};;\mmaarg{n}]]} returns the sublist spanning indices \mmaarg{m} to \mmaarg{n}.
The shortcut \mmafunc{ex /.\mmaarg{a} -> \mmaarg{b}} replaces sub-expressions of \mmafunc{ex} which match pattern \mmaarg{a}, with \mmaarg{b}.
Comments appear in parentheses \mmacomment{(* like this *)}.

Finally, for those copying code into Mathematica, subscripts can be quickly entered into a notebook with keyboard shortcut \keystroke{Ctrl} \& \keystroke{-}.

In the code snippets featured in this manuscript, context should make clear what is \textit{input} to the Mathematica notebook, and what is a rendered \textit{output}.


\subsection{\QuESTlink overview}
\label{sec:how_to_use}

The \QuESTlink package requires no installation, and can be downloaded directly from within Mathematica:

\snippet{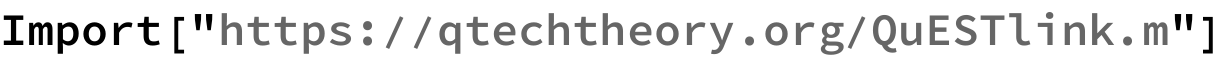}

The user then has a choice of several ``QuEST environments" in which to perform quantum emulation, all of which provide an identical user experience.

\snippet{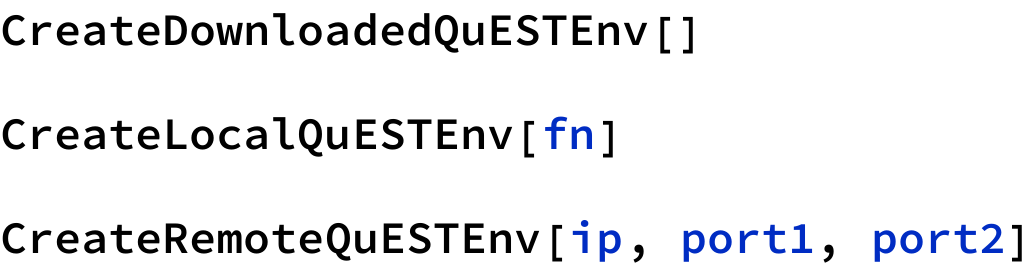}

The first, \mmafunc{CreateDownloadedQuESTEnv[]}, enables simulation on the user's machine by downloading a serial \QuESTlink executable, pre-compiled for the user's operating system. This requires no apriori setup whatsoever.

\mmafunc{CreateLocalQuESTEnv[\mmaarg{fn}]} will attempt to launch an existing local \QuESTlink executable, located at \mmaarg{fn}. This allows local simulation using serial, multi-core or GPU resources, depending on how the executable was compiled. Users can compile \QuESTlink for their platform using the tools on the Github repo~\cite{questlink_github}.

\mmafunc{CreateRemoteQuESTEnv[\mmaarg{ip}, \mmaarg{port1}, \mmaarg{port2}]} connects to a remote \QuESTlink environment at the given \mmaarg{ip} address and ports. 
The remote machine can use serial, multithreading or GPU-acceleration to emulate quantum systems. The facilities to setup a remote \QuESTlink server are provided in the Github repo~\cite{questlink_github}.

Once connected to a \QuEST environment, a full list of the supported \QuESTlink facilities and gate symbols can be obtained by evaluating

\snippet{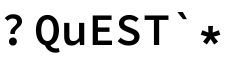}

and the documentation of a particular function or operator obtained similarly using \mmafunc{?}.

\snippet{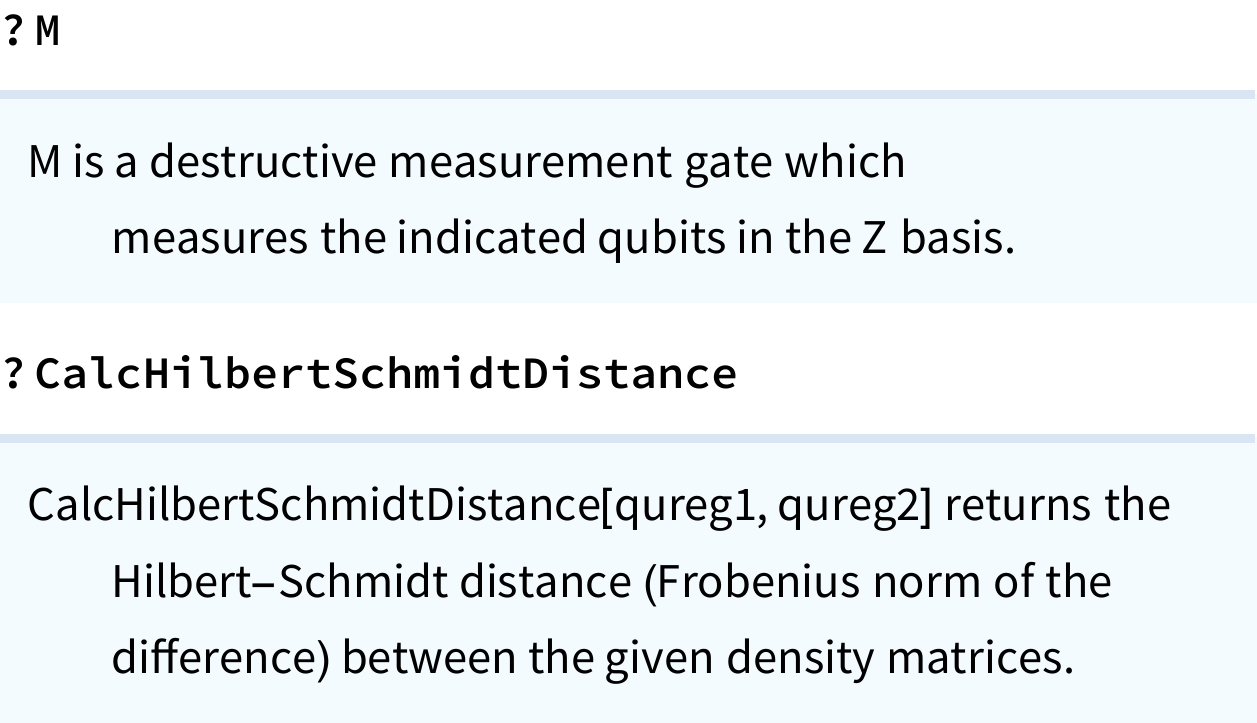}

Emulation begins by creating quantum registers (a ``\mmafunc{Qureg}"), each represented by a state-vector or density-matrix.

\snippet{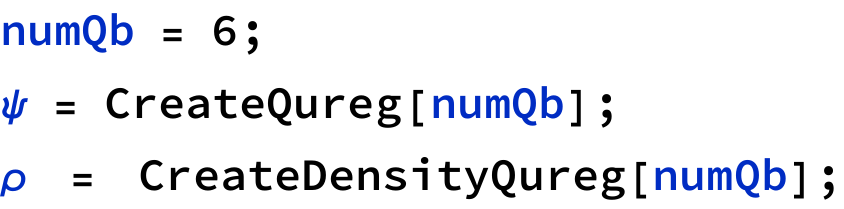}

These functions return a unique ID for each \texttt{Qureg}, the memory for which is stored in the \QuEST environment.
Once created, \QuESTlink provides a few functions to initialise \mmafunc{Qureg}s:

\snippet{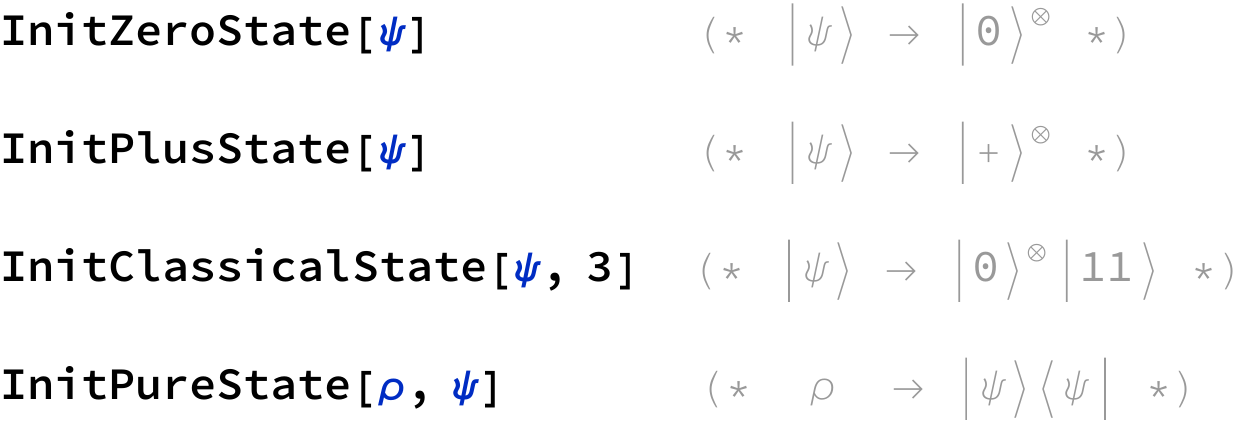}

or directly modify them (here, creating a random density matrix):

\snippet{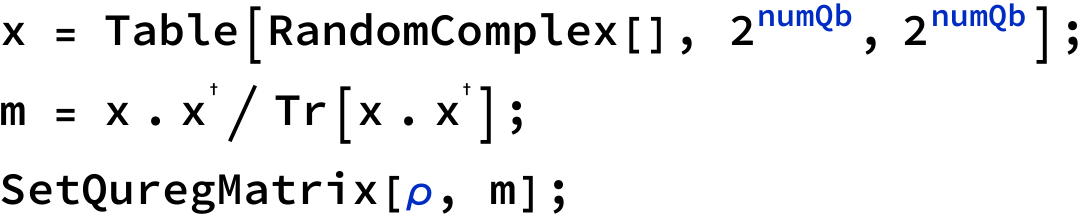}

A quantum circuit $\textcolor{mmablue}{u}$ can be applied to a \mmafunc{Qureg}, agnostic of whether it is a state vector (effecting $ \textcolor{mmablue}{u} \, \ket{\psi} $) or a density matrix (effecting $\textcolor{mmablue}{u} \, \rho \, {\textcolor{mmablue}{u}}^\dagger$), using \mmafunc{ApplyCircuit[\mmaarg{u}, \mmaarg{qureg}]}. \QuESTlink features a concise and expressive language for specifying gates and decoherence processes, where target qubits are denoted with subscript integers to gate symbols, and control gates ``wrap'' the base gate. The below example makes use of the \mmafunc{Circuit} function, which disables commutation and allows a concise product representation of the circuit. In combination, this enables a syntax akin to how circuits are denoted in the quantum computing literature.

\snippet{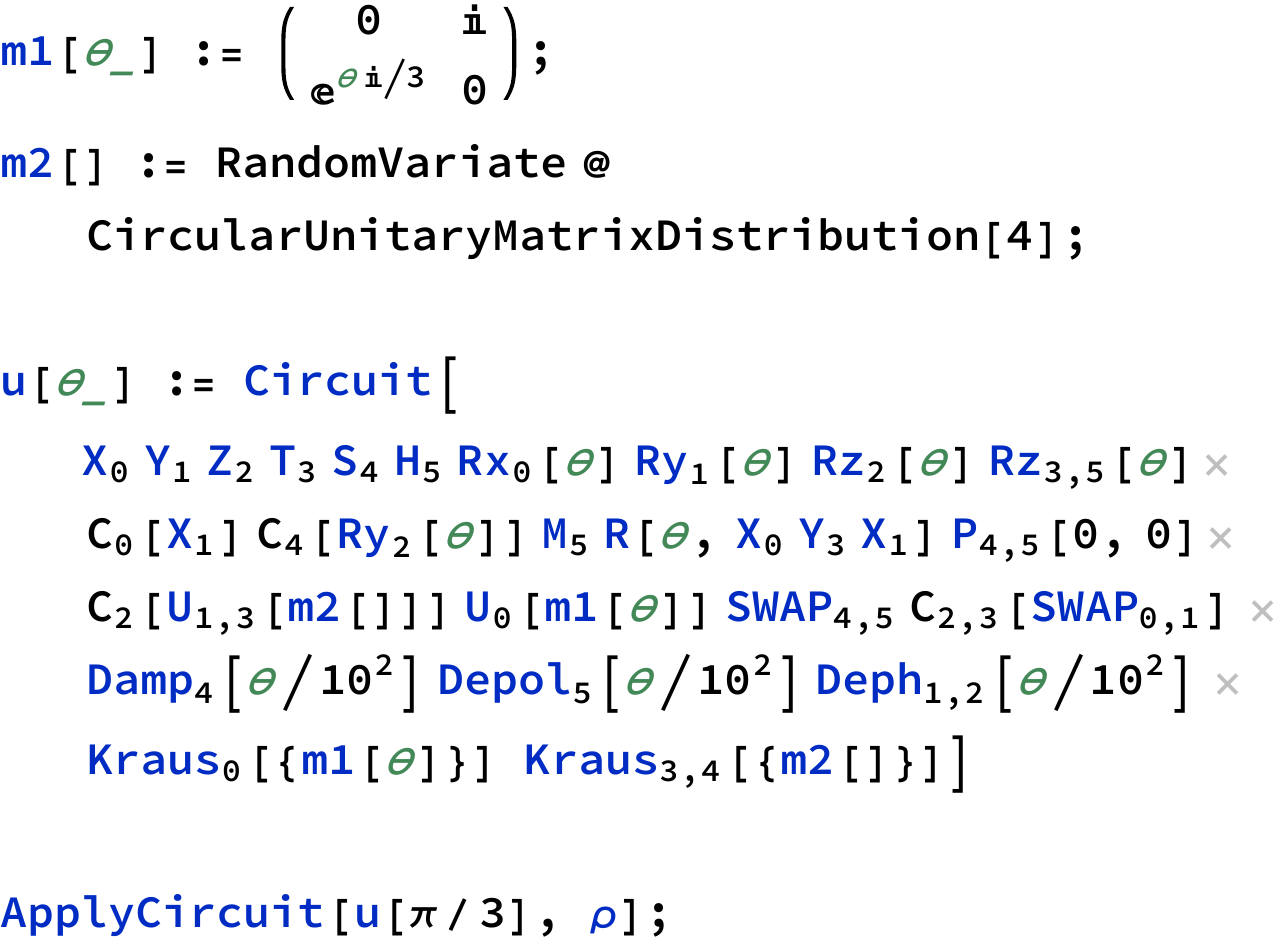}

The result of \mmafunc{Circuit[]} is just a list of operators, allowing easy circuit manipulation. Mathematica features many tools for such lists allowing the user to easily extend, alter, join, compare, etc, their quantum circuit. For example: 

\snippet{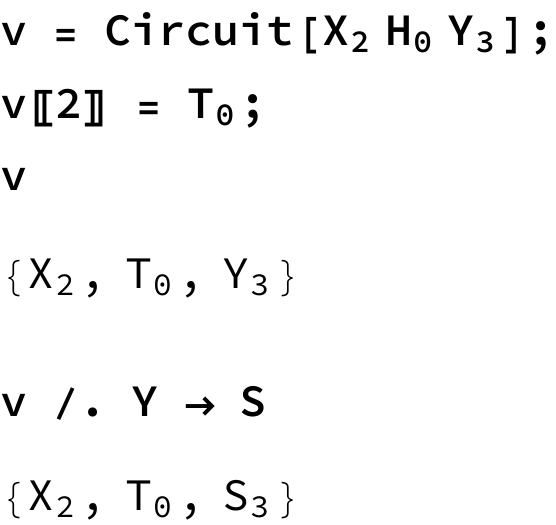}

These circuits can be swiftly visualised with \mmafunc{DrawCircuit[\mmaarg{u}]}, which supports saving rendered circuit diagrams to raster and vector images.

\snippet{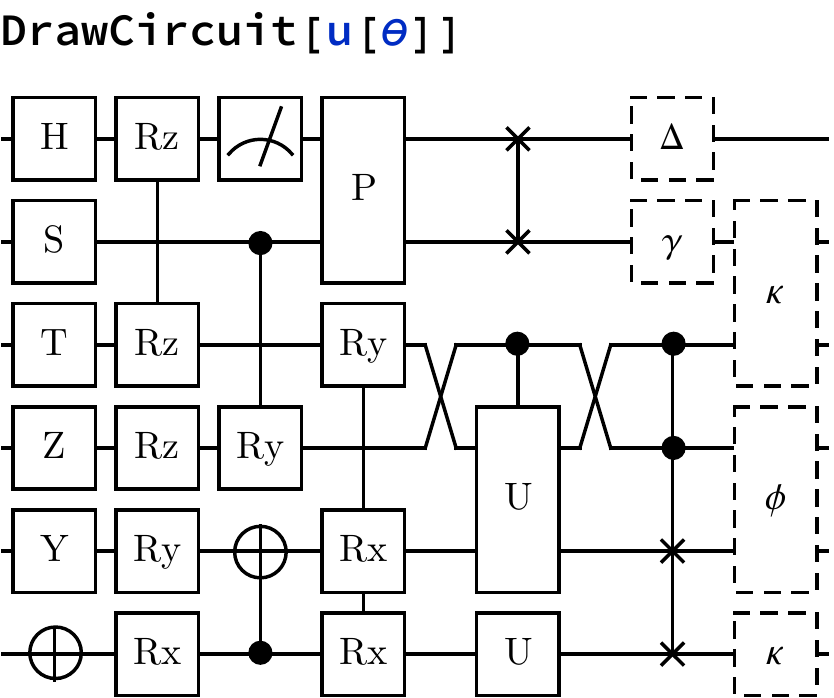}

After applying a unitary circuit, one can of course opt to obtain the entire state vector or density matrix, as for example: 

\snippet{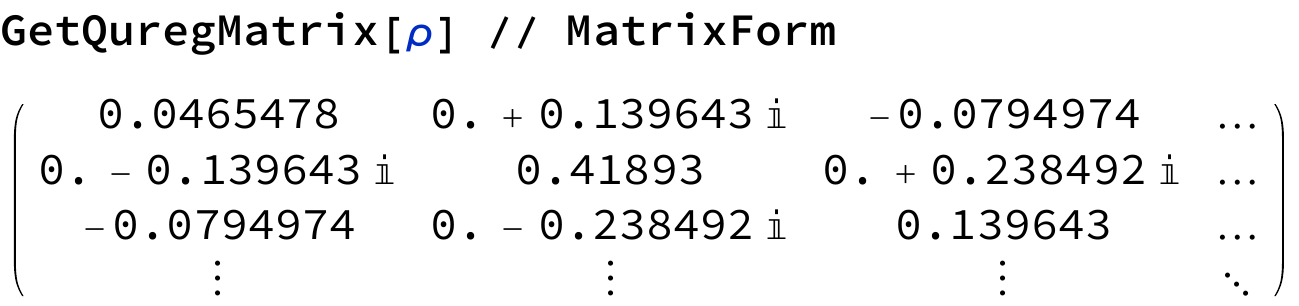}

However, it is typically more efficient to calculate desired quantities, like the expectation value of a Hamiltonian in the Pauli-basis, in the \QuEST environment itself.

\snippet{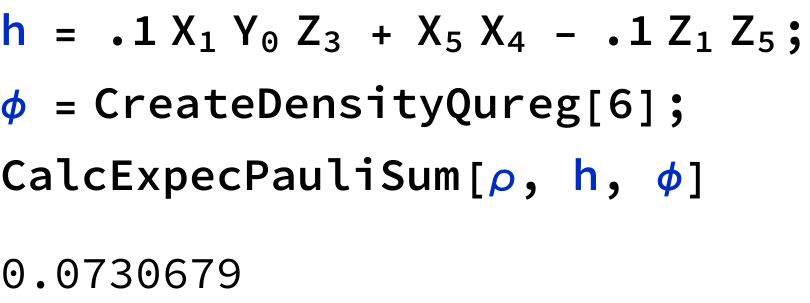}

At any time, \mmafunc{Qureg}s can be individually (or, all-together) destroyed to free up memory in the \QuEST environment.

\snippet{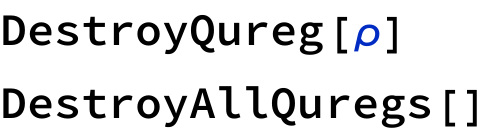}

By leveraging Mathematica's powerful suite of symbolic calculations, {\QuESTlink} can even build analytic matrices from circuit specifications.

\snippet{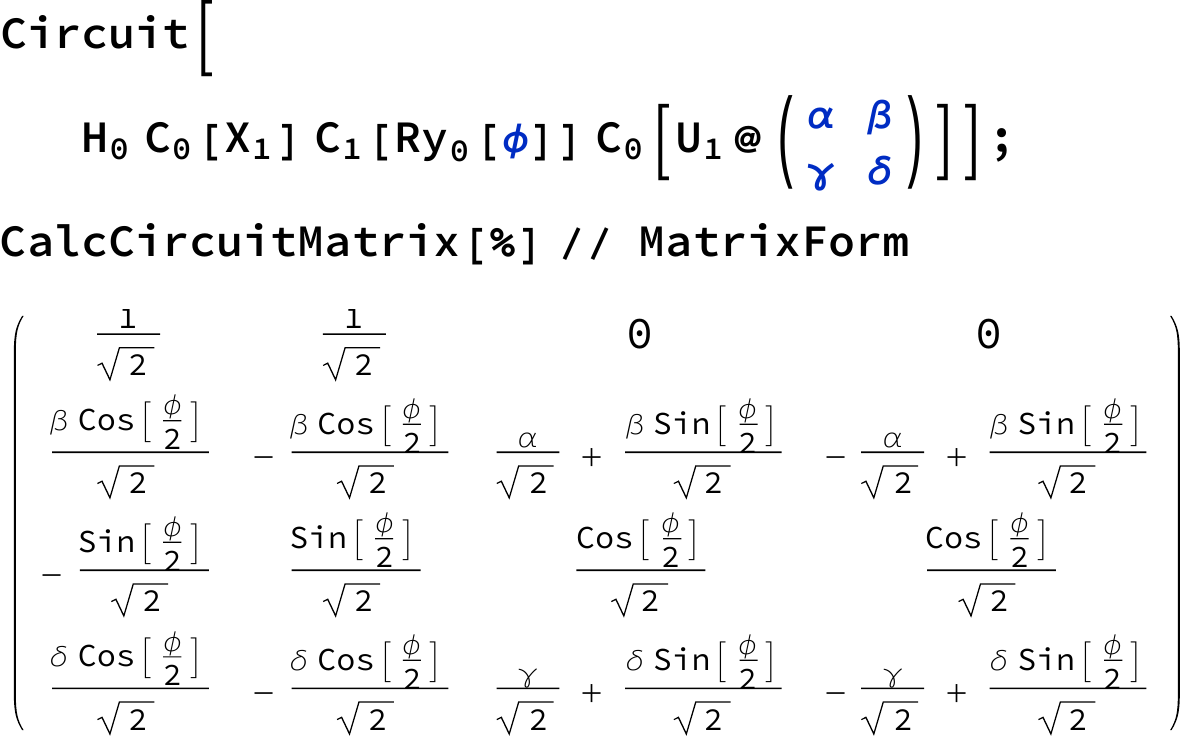}

With these facilities, \QuESTlink offers a seamless integration with Mathematica's comprehensive range of computational and graphical tools, as illustrated in the following demonstrations.


\section{Demonstrations}

To demonstrate the concision~\cite{concision_newyorker} offered by \QuESTlink, we provide several examples of somewhat sophisticated computations written in only several lines, and efficiently simulated. For users wishing to run these demonstrations directly, they are compiled into a single notebook at \href{https://questlink.qtechtheory.org/paper_demos.nb}{\url{questlink.qtechtheory.org/paper_demos.nb}}.

\subsection{Decoherence}

To begin, we very compactly demonstrate the effect of two-qubit depolarising noise on the expected measurement of a simple Hamiltonian \mmaarg{h}. Starting in a random pure state $\textcolor{mmablue}{\psi}$, a depolarising channel with probability $0.1$ of any Pauli error occurring is repeatedly applied, in total $100$ times.

\snippet{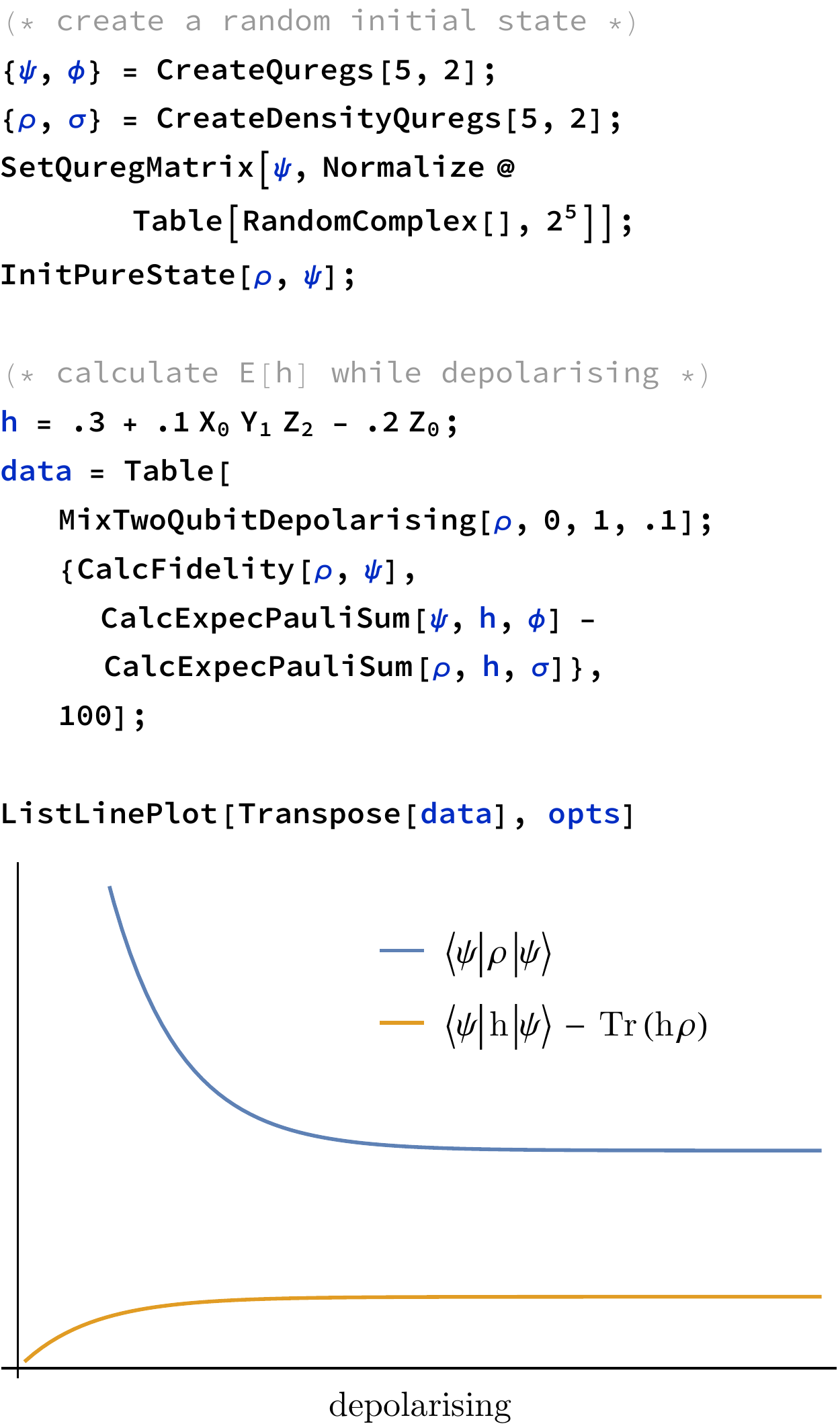}

Note that undisclosed variable \mmaarg{opts} contained additional code for customising the plot.

\subsection{Variational imaginary-time}

In this demonstration, we emulate the quantum variational imaginary-time simulation routine~\cite{McArdle_imagtime} to approximate the ground-state of a molecular Hamiltonian.

We first download a reduced 6-qubit representation of the electronic structure Hamiltonian of Lithium Hydride (LiH).

\snippet{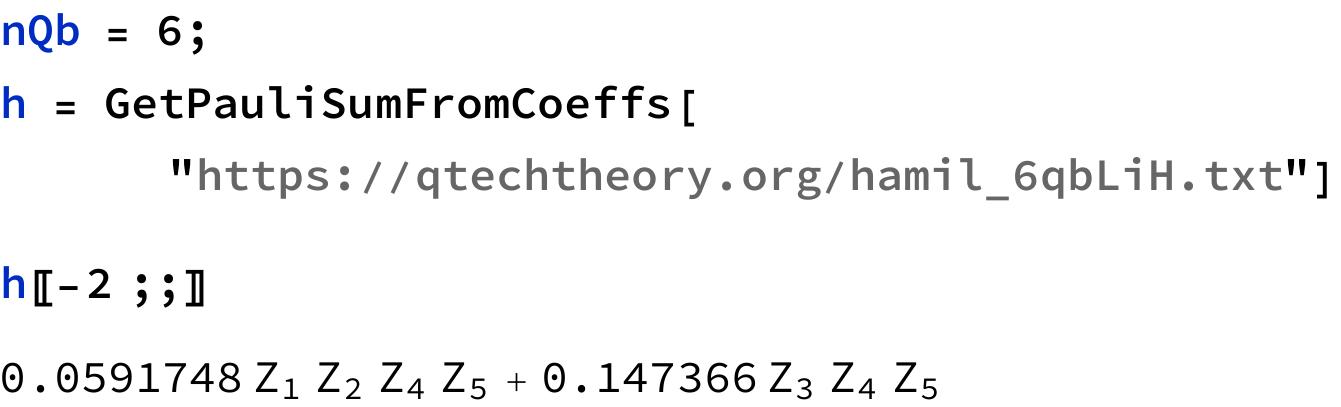}

We create a simple 6-qubit ansatz circuit featuring 39 parameters, denoted with variables $\vec\theta = \{\textcolor{mmablue}{\theta_1}, \dots, \textcolor{mmablue}{\theta_{39}}\}$.

\snippet{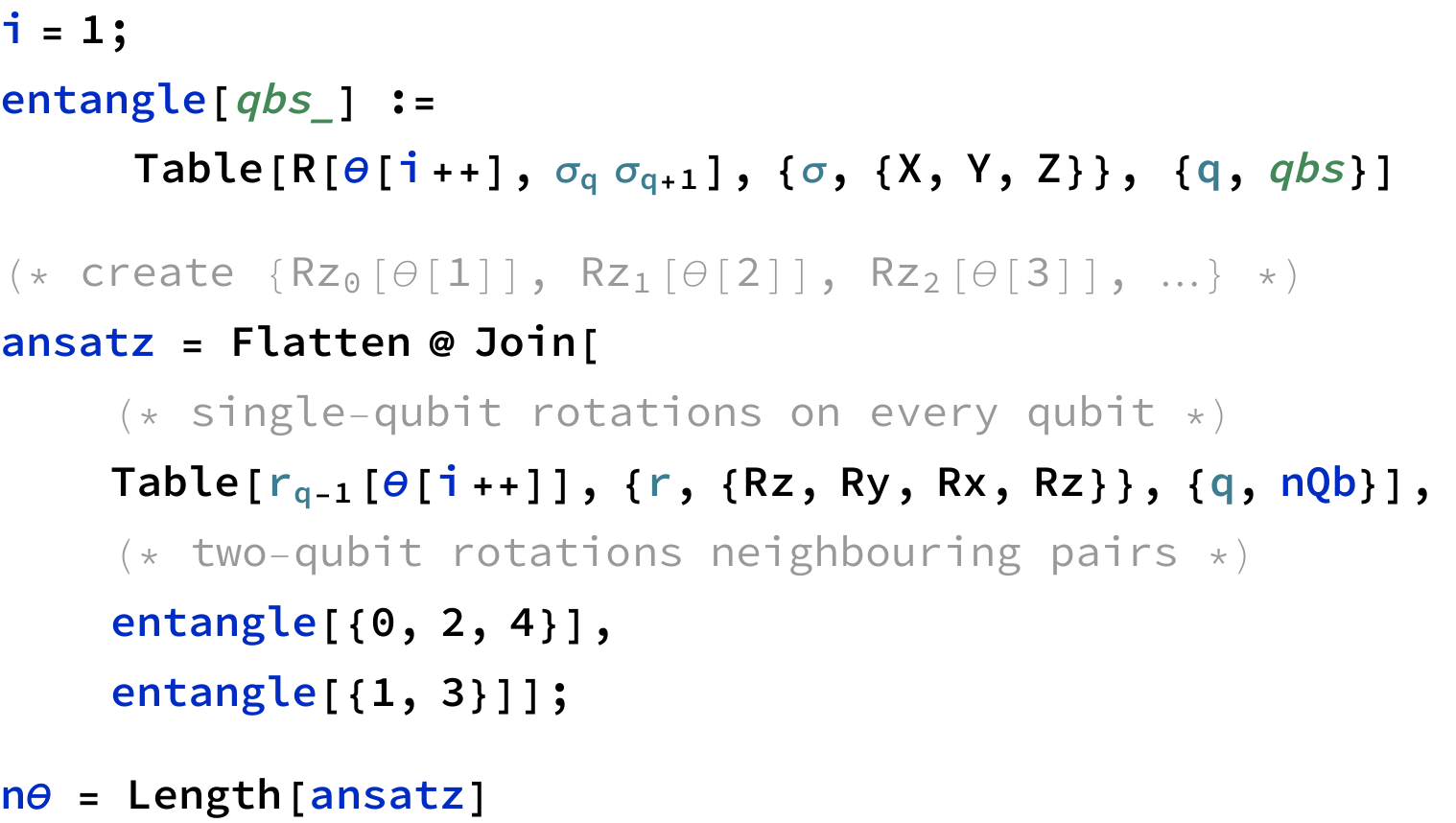}

This \mmaarg{ansatz} circuit consists of one and two qubit rotation gates; $\exp(-i \textcolor{mmablue}{\theta_j} \hat\sigma/2)$, for $\hat\sigma \in \{X,Y,Z,X\otimes X,Y\otimes Y,Z\otimes Z\}$, which in \QuESTlink are denoted by $\mmafunc{Rx}_\mmaarg{q}\mmafunc{[}\textcolor{mmablue}{\theta}\mmafunc{]}$ (etc.) and $\mmafunc{R[}\textcolor{mmablue}{\theta}\mmafunc{,X}_\mmaarg{q1}\mmafunc{X}_\mmaarg{q2}\mmafunc{]}$. In the diagram below, each such paired two-qubit rotation is marked by a vertical link.

\snippet{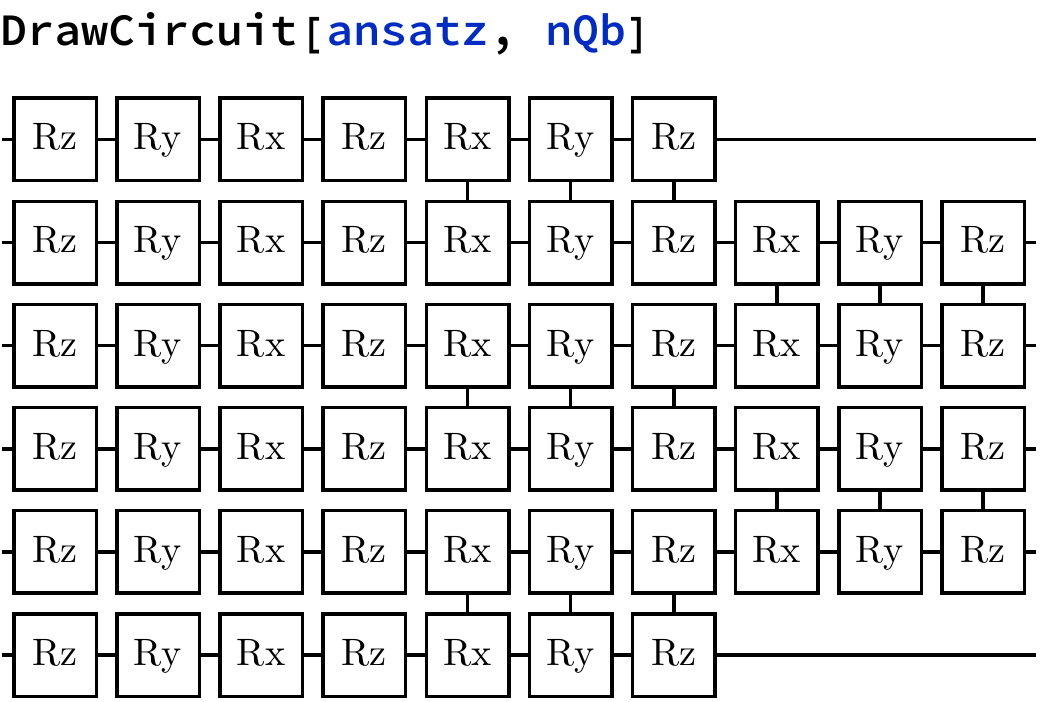}

We now create several \mmafunc{Qureg}s with which to emulate the quantum algorithm.

\snippet{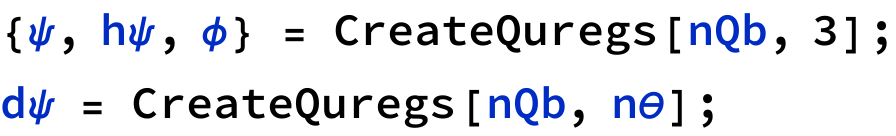}

\mmaarg{$\psi$} will be maintained as the output state of the ansatz circuit, and \mmaarg{$h\psi$} will store the result of applying the Hamiltonian \mmaarg{$h$} to \mmaarg{$\psi$}. \mmaarg{$\phi$} will merely provide intermediate work-space for calculations, and \mmaarg{$d\psi$} will store a \mmafunc{Qureg} for each parameter in the ansatz (a total of $\textcolor{mmablue}{n\theta}$).

Next, we choose a random initial assignment of the ansatz parameters, and measure the energy of the resulting quantum state.

\snippet{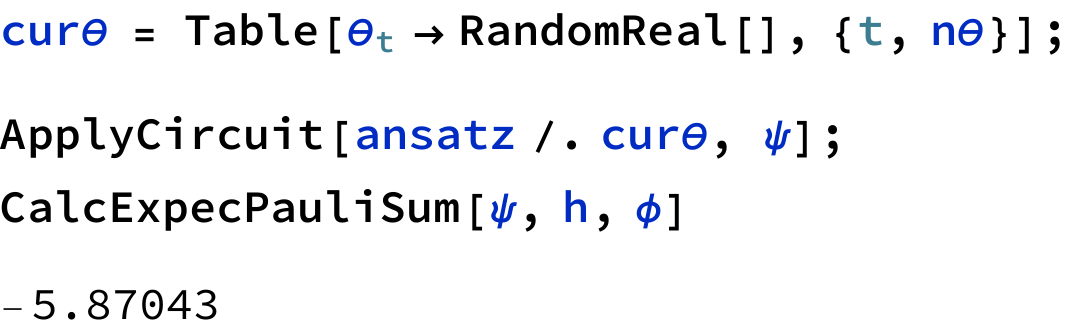}

The variational imaginary-time algorithm~\cite{McArdle_imagtime} involves repeatedly measuring a matrix and vector of quantities,
\begin{align}
    A_{ij} &= \text{Re}\left\{
        \bigl< \frac{\partial \textcolor{mmablue}{\psi}(\vec\theta)}{\partial \theta_i} | \frac{\partial \textcolor{mmablue}{\psi}(\vec\theta)}{\partial \theta_j} 
        \bigr>
    \right\},
    \\
    C_i &= - \text{Re}\left\{
    \bigl< \textcolor{mmablue}{\psi}(\vec\theta) | \textcolor{mmablue}{h} | \frac{\partial \textcolor{mmablue}{\psi}(\vec\theta)}{\partial \theta_i} \bigr>
    \right\}
\end{align}
and iteratively updating the parameters under
\begin{align}
    \vec\theta \to \vec\theta + \textcolor{mmablue}{\Delta t} \, \textcolor{mmablue}{A}^{-1} \, \textcolor{mmablue}{\vec{C}},
\end{align}
which we now concisely emulate for \mmaarg{nt} iterations.

\snippet{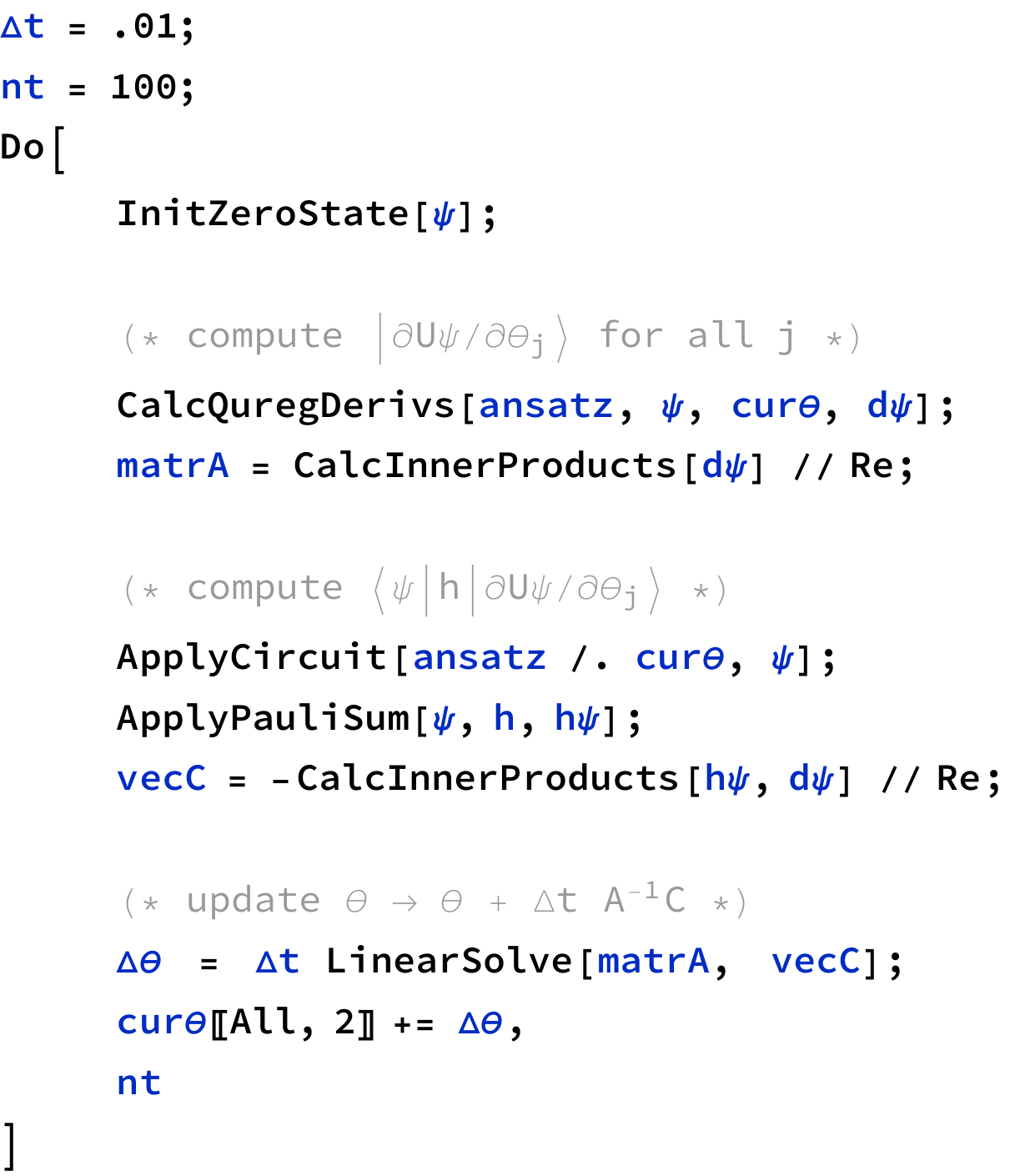}

The energy of the output quantum state, as produced by the reached assignment of the parameters, is close to the true groundstate found through matrix diagonalisation.

\snippet{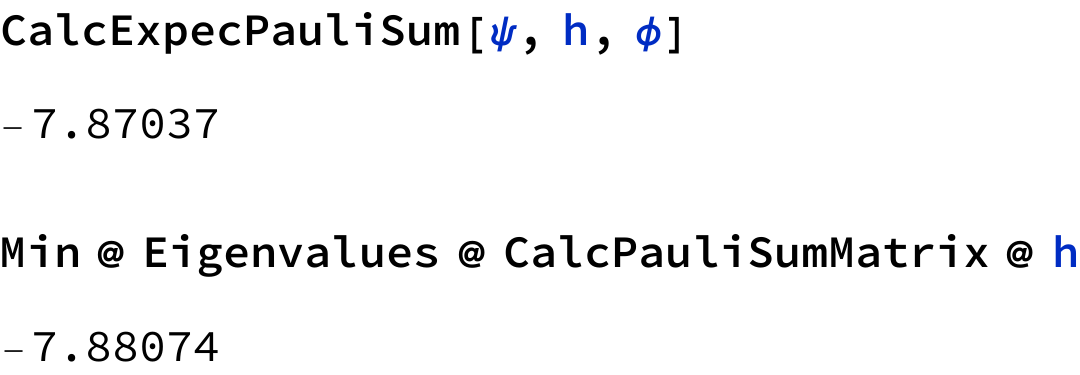}

Interestingly, a significantly slower but direct minimisation in Mathematica reveals the energy reached by imaginary-time was \textit{not} quite the best possible of our chosen ansatz.

\snippet{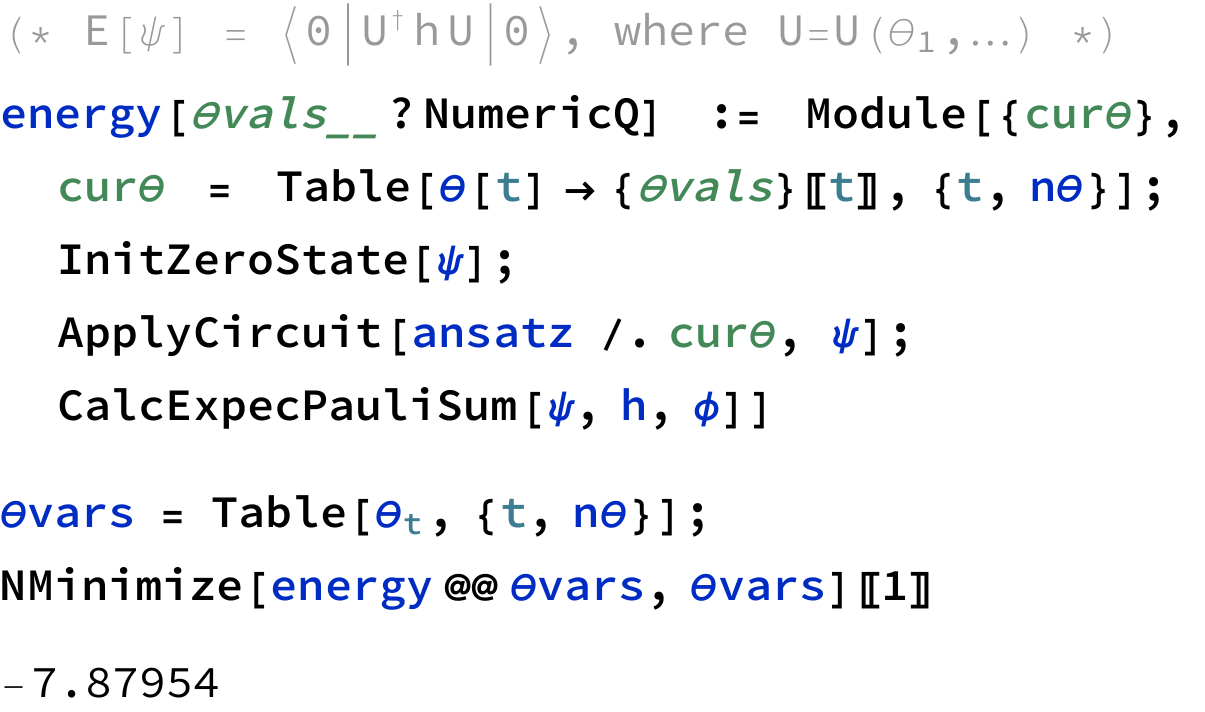}

We invite the interested reader to compare the Mathematica code above to a native C implementation of imaginary time evolution, as hosted on Github~\cite{imagtime_github}.

\subsection{Noisy Trotterisation}

In this demonstration, we emulate Trotterisation of a spin-ring Hamiltonian, with and without noise, and compare it to direct numerical solving of the Schr\"odinger equation.
Our Hamiltonian is the one-dimensional nearest-neighbour (periodic boundary conditions) \mmaarg{nQb}-spin Heisenberg model with a random magnetic field in the $z$ direction, 
\begin{align}
    \sum\limits_j^{\mmaarg{nQb}} \vec\sigma_j \vec\sigma_{j+1} + r_j \sigma_j^z,
    \;\;
    \;\;
    r_j \sim \mathcal{U}[-1,1].
\end{align}
Real-time simulation of this model to time $t=\mmaarg{nQb}$ is nominated by Childs~\textit{et al.}~\cite{Childs_rand_hamil} as an early practical application of a quantum computer. In this example, we'll study a five-spin chain Hamiltonian \mmaarg{h}.

\snippet{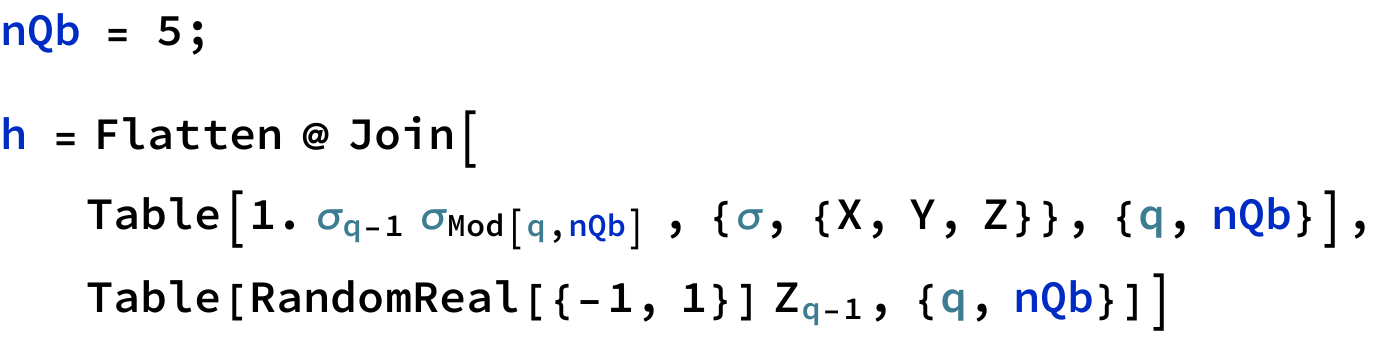}

To simulate evolution on a quantum computer, we will emulate circuits formed by the Suzuki-Trotter decompositions~\cite{Suzuki_trotter} of the unitary evolution operator of varying order \mmaarg{n}. Below, \mmaarg{r} is the number of repetitions of the order-\mmaarg{n} circuit to perform, to ultimately reach time \mmaarg{t}.

\snippet{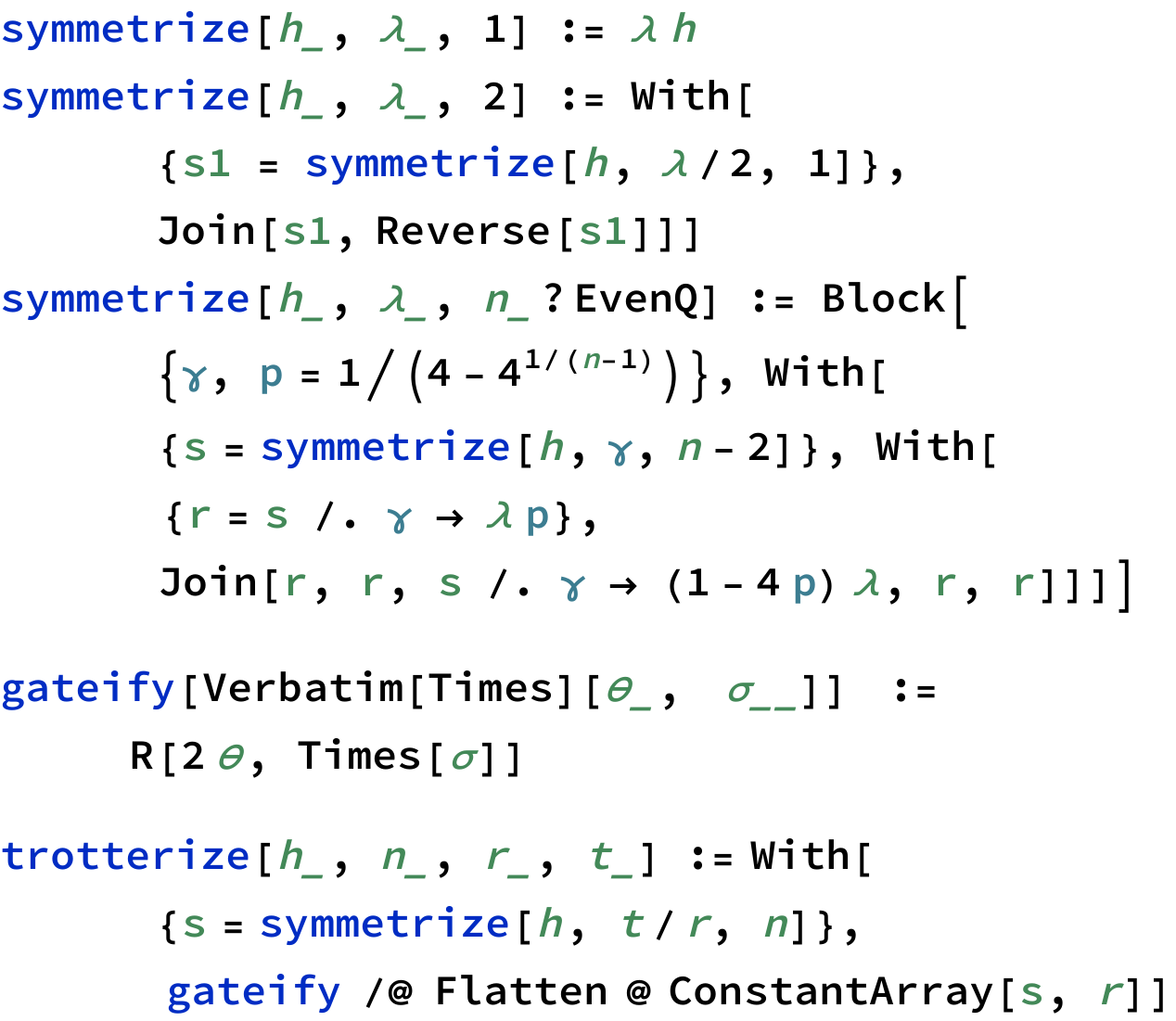}

Note the original ordering of the Hamiltonian terms is arbitrary, and should ideally be optimised into commuting groups, or at least randomly shuffled. In particular, Childs~\textit{et al.}'s uniform randomisation scheme~\cite{Childs_rand_trotter}, whereby each repetition sees a random Hamiltonian ordering, is trivially implemented as:

\snippet{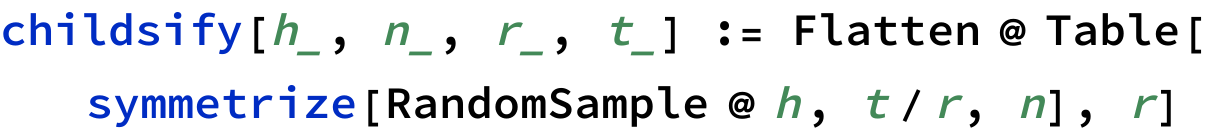}

Even Campbell's qDRIFT routine~\cite{Campbell_qdrift} can be given a compact implementation.

\snippet{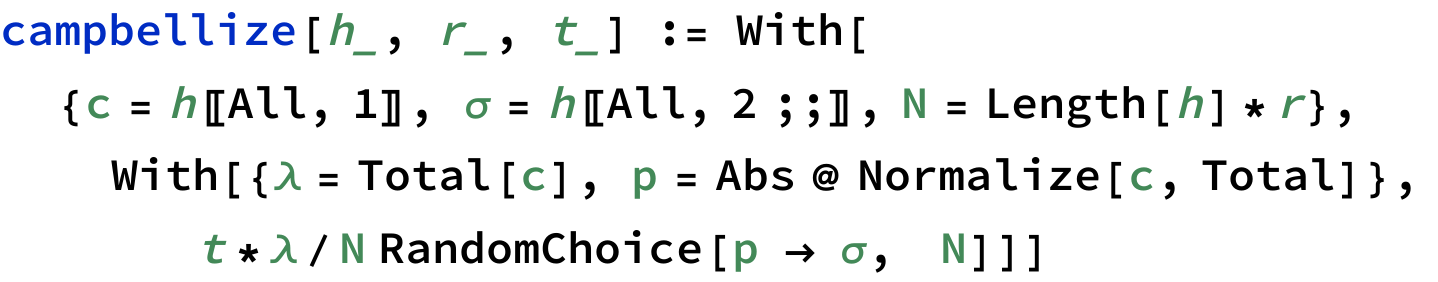}

However, we opt instead to utilise only deterministic Trotterization for clarity. For example, the first-order (\mmaarg{n}\mmafunc{=1}) single-repetition (\mmaarg{r}\mmafunc{=1}) Trotter circuit (arbitrarily to time \mmaarg{t}\mmafunc{=1}) has the form: 

\snippet{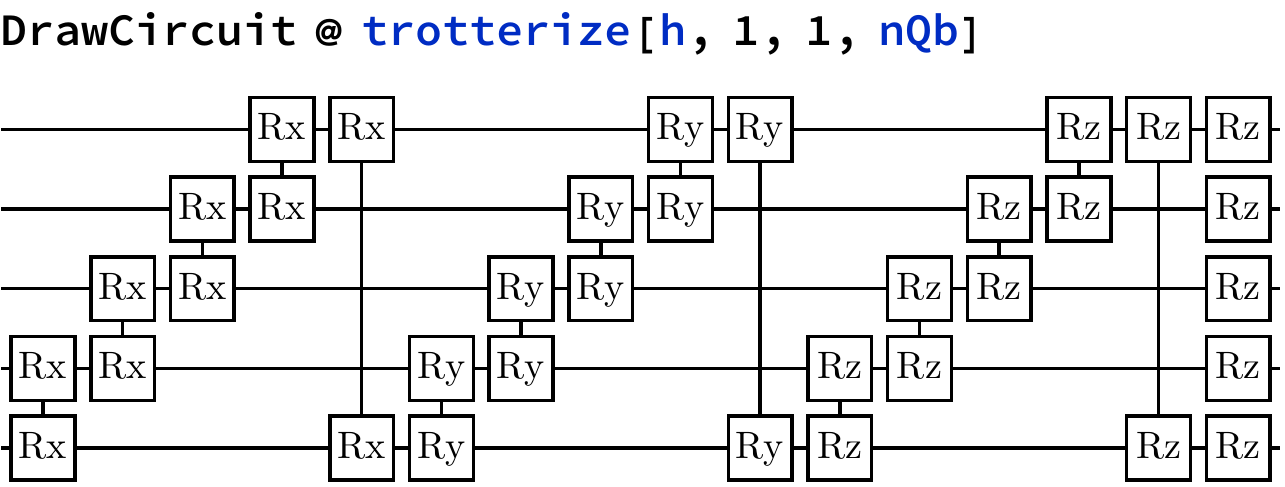}

We compare the states produced from Trotter circuits with the ``true" (to numerical precision) time evolution, found by numerically solving the Schr\"odinger equation using Mathematica's in-built \mmafunc{NDSolveValue} routine.

\snippet{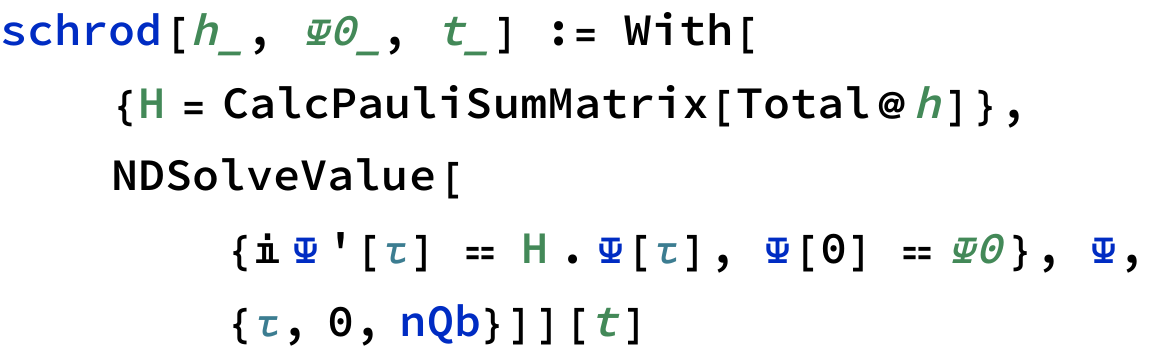}

In the proceeding solutions, the \mmafunc{Qureg} $\textcolor{mmablue}{\psi0}$ will store the initial state (a random pure state), $\textcolor{mmablue}{\psi t}$ will store the true state, and $\textcolor{mmablue}{\psi}$ will store outputs of the Trotter circuits.

\snippet{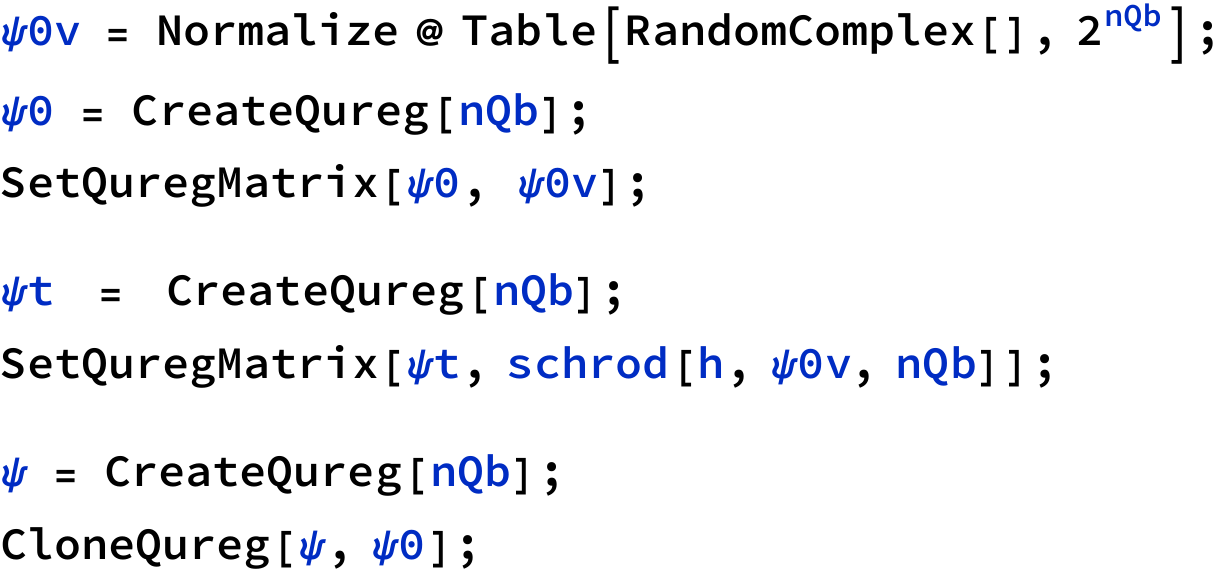}

We can now commence the computation by generating circuits of a varying Trotter order and number of repetitions, emulating their execution, and computing the fidelity of their output state with the true state.

\snippet{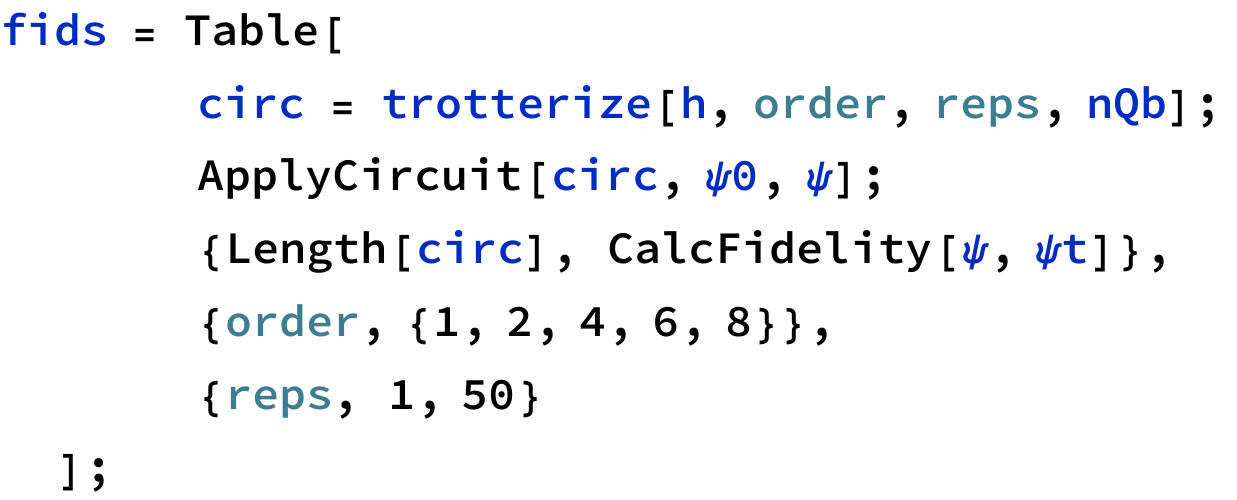}

We then plot the result below, using some undisclosed parameters \mmaarg{opts} to tweak the plot style. The horizontal axis is the total number of gates in the Trotter circuit.

\snippet{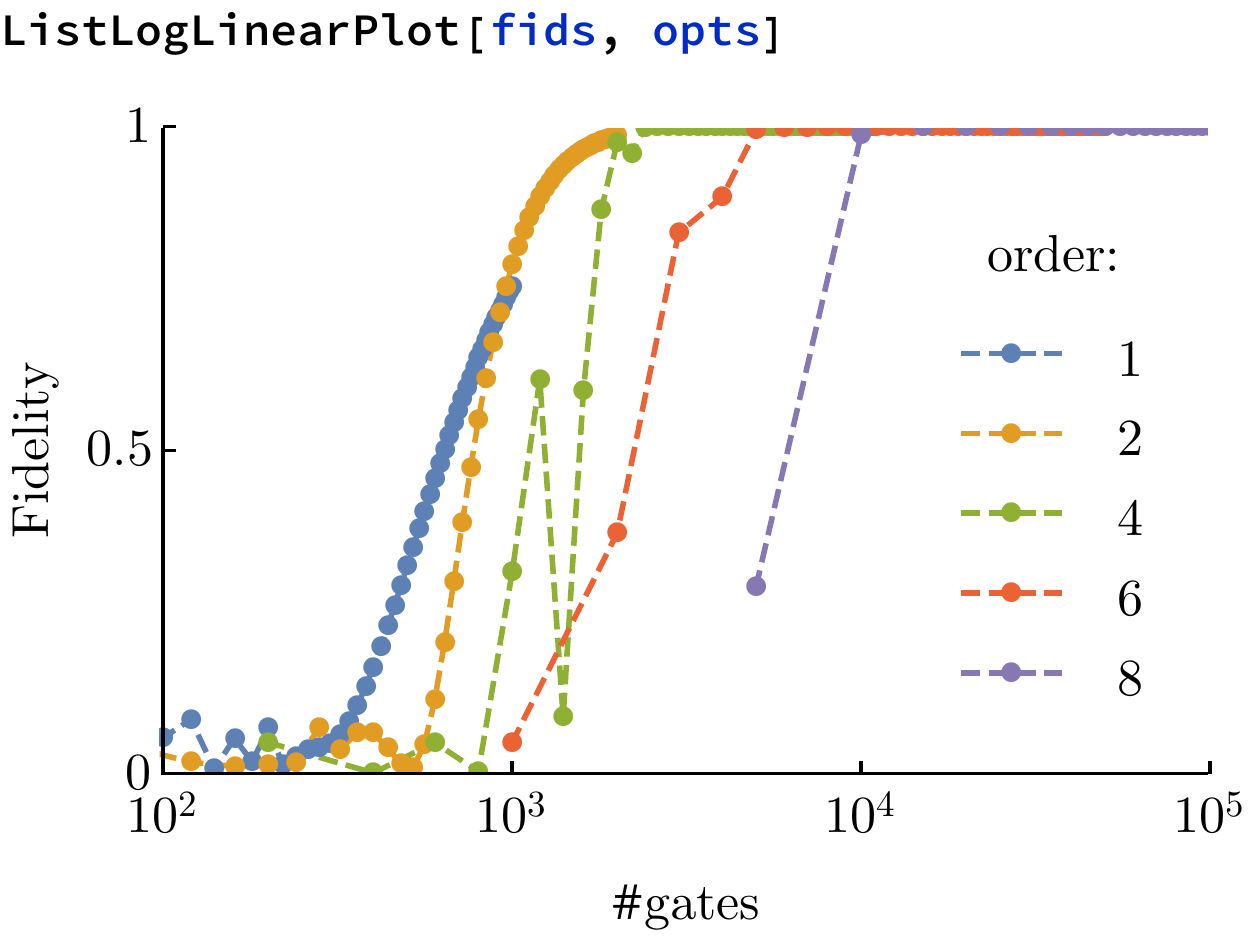}

We can even imitate an noisy quantum device by inserting decoherence operators into our circuit. In this demonstration, we will follow each Trotter-prescribed gate with one or two qubit depolarising operators, which effect the channel
\begin{align}
    \mmafunc{Depol}[\textcolor{mmablue}{\rho}] = (1 - p)\textcolor{mmablue}{\rho} + \frac{p}{3} \sum\limits_{\sigma} \sigma \textcolor{mmablue}{\rho} \sigma,
\end{align}
(noting $\sigma=\sigma^\dagger$ for Paulis) and similarly for the two-qubit analogue.
We choose an error rate of $p=10^{-4}$, below the rates recently demonstrated in Google's Sycamore machine
~\cite{Google_supremacy}. Below is a visualisation of the first five unitary gates of the resulting noisy circuit.

\snippet{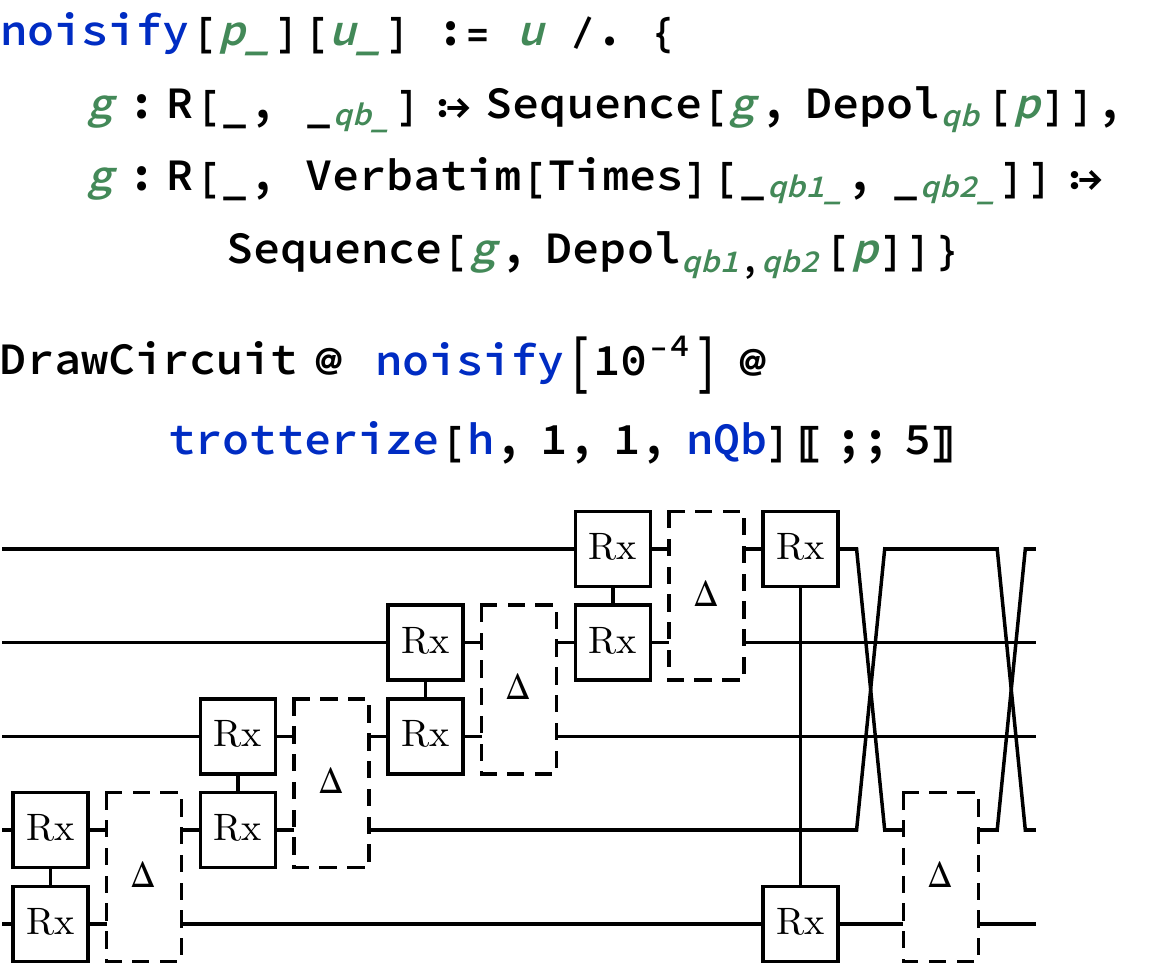}

Our simulation of the noisy circuit is near-identical to our simulation of pure states, with the exception that we now operate upon density matrices.

\snippet{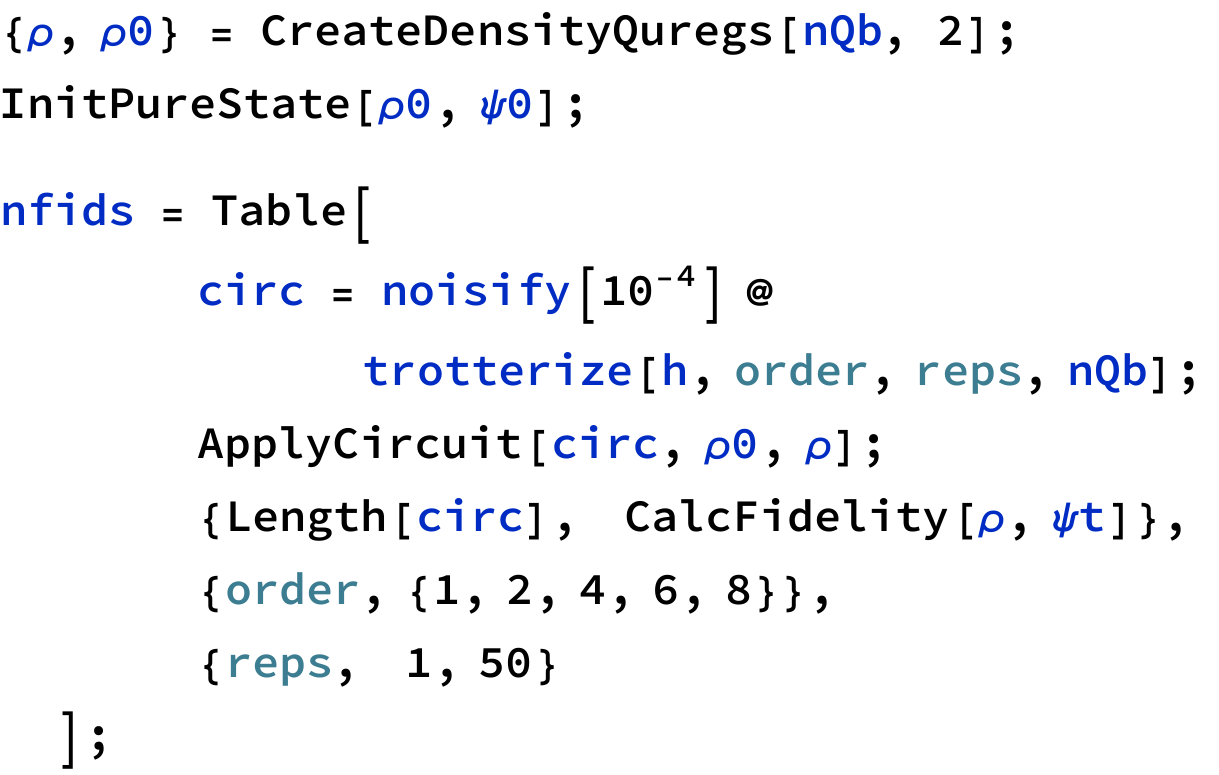}

As one might expect, the accuracy of the Trotter circuits have waned, and eventually decrease with increased circuit depth, due to the opportunity for additional errors to accrue.

\snippet{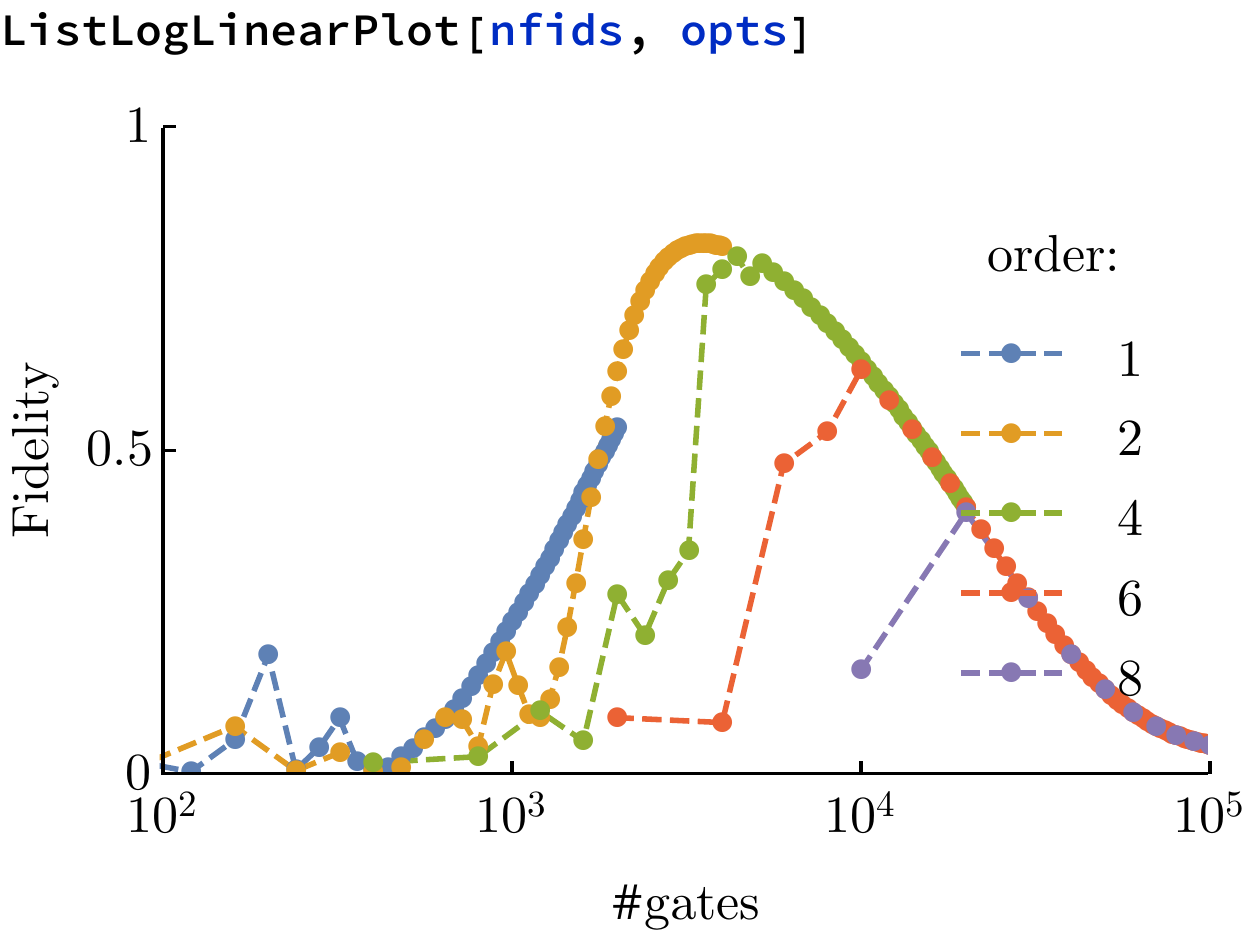}

\subsection{Further examples}

Complete notebooks demonstrating some extended computations in \QuESTlink are available at \href{https://questlink.qtechtheory.org/}{questlink.qtechtheory.org}.


\section{Benchmarking}
\label{sec:demonstrations}

We now benchmark \QuESTlink's local and remote emulation of some arbitrary quantum circuits, and compare to benchmarks of direct simulation in \texttt{C}, using \QuEST. This quantifies the runtime overheads incurred by Mathematica integration, and whether \QuESTlink is the right emulation tool for the user's target simulation regime.

\begin{figure}
    \centering
    \includegraphics[width=.48\textwidth]{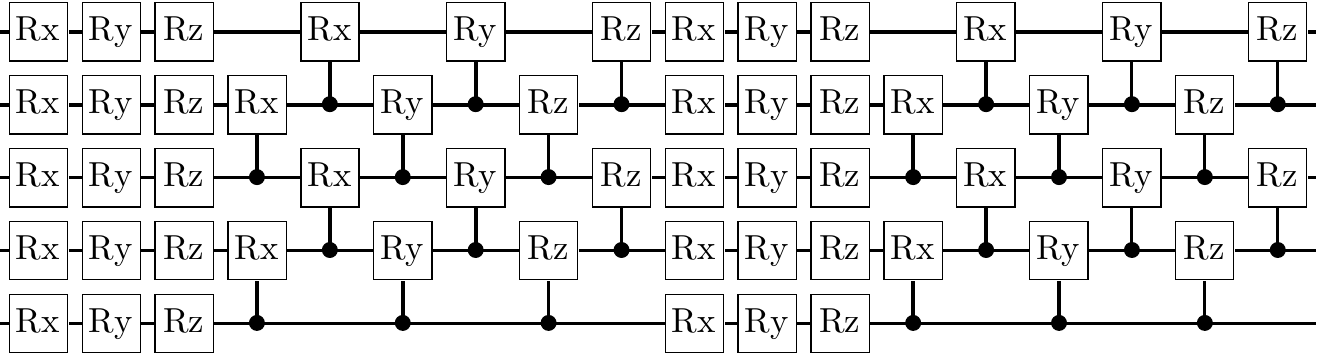}
    \caption{A $5$-qubit $2$-repetition example of the arbitrary circuit used for comparative benchmarking of \QuESTlink and \QuEST. Every gate involves a rotation of a uniformly random angle between $0$ and $4\pi$, which generate all rotations. The $15$-qubit circuits employed for benchmarking feature 87 gates per repetition.}
    \label{fig:benchmark_circ}
\end{figure}

We nominate a simple circuit consisting of $X$, $Y$, and $Z$ axis rotations of random angles on each qubit, followed by tessellated controlled rotations on every pair of neighbouring qubits. This pattern is repeated in the circuit, and is illustrated in Figure~\ref{fig:benchmark_circ}. These benchmarks will emulate $15$-qubit circuits, with between $1$ to $50$ repetitions (an upper bound of 4350 gates), and each will be simulated $10$ times with re-randomised angles.

Core \QuEST is profiled directly in \texttt{C}, through precise timing of each full circuit execution. In contrast, \QuESTlink is profiled through a notebook using Mathematica's \mmafunc{AbsoluteTiming[]} function; its runtime will include Mathematica evaluation, \texttt{QuESTlink.m} preprocessing and circuit encoding, the \texttt{C/Link} and WSTP overheads, \texttt{quest\_link.c}'s decoding of the circuit, and ultimately \QuEST's simulation.

Benchmarking is performed on a 12-core Xeon W-2133 3.6\,GHz CPU. At most 8 threads will be employed in multithreaded mode, so as not to interfere with threads used by the Mathematica kernel. GPU-accelerated testing will employ a 24\,GB NVIDIA Quadro P6000 in the same machine.

\begin{figure}
    \centering
    \makebox[.1\textwidth][c]{\includegraphics[width=.5\textwidth,trim={.2cm 0 0 0},clip]{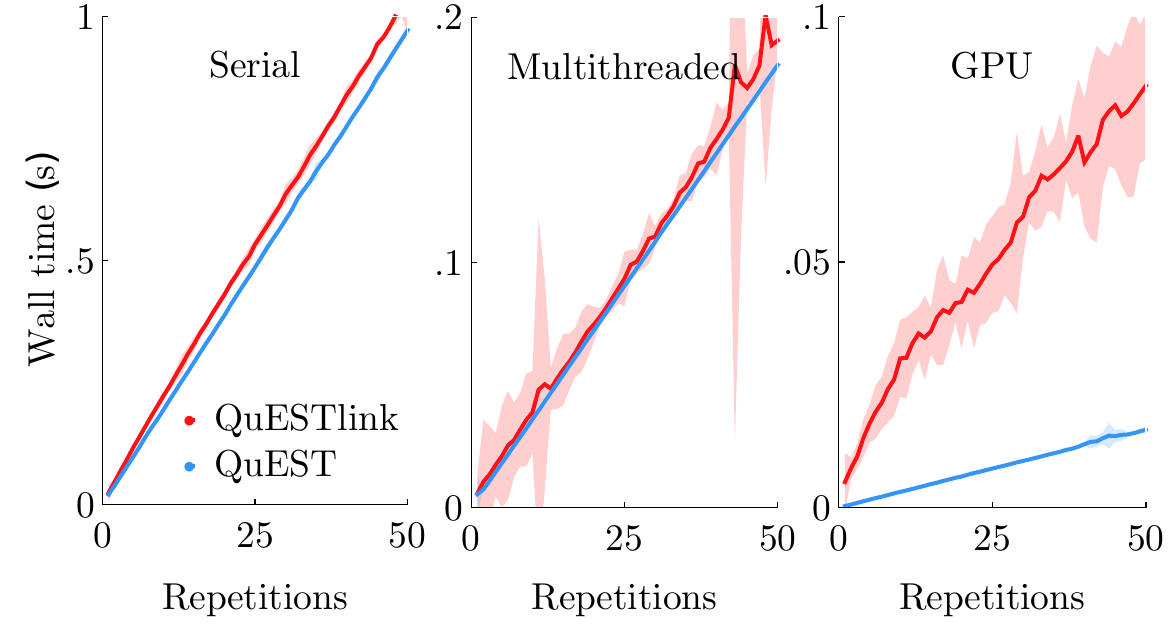}}
    \caption{Comparitive performance of \QuEST and (local) \QuESTlink emulating the 15-qubit circuit illustrated in Figure~\ref{fig:benchmark_circ}, with a varying circuit depth (repetitions of the base circuit), in each of \QuESTlink's supported parallelisation modes. Multithreaded emulation used $8$ of $12$ available threads, though were shared with the Mathematica kernel.
    Solid lines indicate the mean runtime of 10 random simulations, and shaded regions indicate three standard deviations. The circuit of $50$ repetitions contains a total of 4350 gates.
    }
    \label{fig:benchmark_results}
\end{figure}

The results of benchmarking are presented in Figure~\ref{fig:benchmark_results}, and are as expected. The Mathematica overhead of invoking serial \QuEST through \QuESTlink is small (a factor $\approx 1.1$). 
This manifests from a fixed per-gate cost of encoding the circuit, though its proportion of the total runtime will depend on the total runtime, and hence the number of simulated qubits.

Multithreading with 8-threads offered a $\approx 5\times$ speedup but
introduced additional variation in runtime, most likely due to dynamic reallocation of threads between the Mathematica kernel and the backend \QuEST process. Otherwise, the multithreading overhead is similarly small (a factor $\approx 1.08$ for these tests). This reveals that the Mathematica pre-processing itself is accelerated by multithreading with the remaining four available threads.

At first glance, GPU \QuESTlink may appear anomalously slow; on average $7.2$ times slower than core GPU \QuEST.
However, this is due to QuEST simulation being very rapid, yielding a proportionately larger Mathematica overhead.
For example, in the largest serial simulation, \QuESTlink was on average $5\times10^{-2}$\ seconds slower than \QuEST. This is comparable to the largest overhead of $4\times10^{-2}$ in the GPU tests.

\begin{figure}
    \centering
    \includegraphics[width=.47\textwidth]{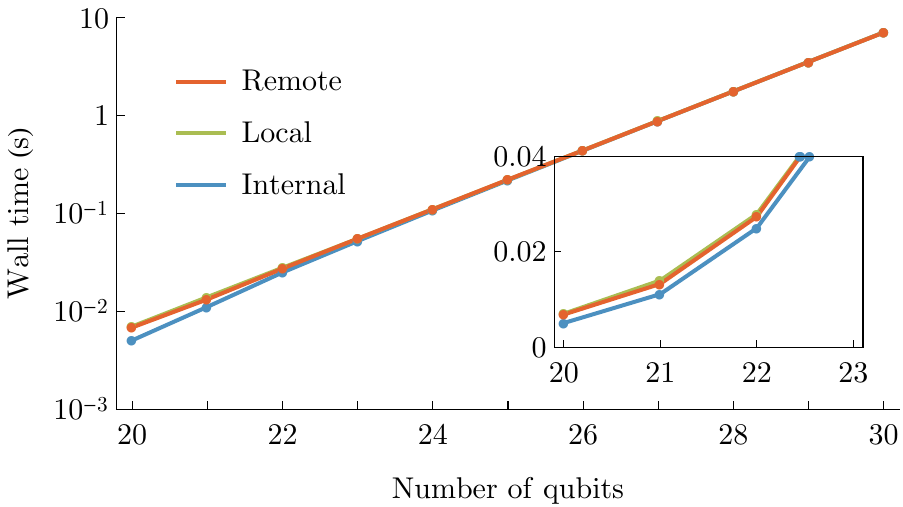}
    \caption{Benchmarking of a fixed-length circuit simulation in QuESTlink, using different environment configurations, though all ultimately running on the previously profiled Xeon machine's GPU. \textit{Remote} used \texttt{CreateRemoteQuESTEnv[]} to connect from a 13-inch 2017 MacBook Pro, with an average latency of $0.38\,$ms, with wall time measured in Mathematica with \texttt{AbsoluteTiming[]}. \textit{Local} ran in a notebook on the Xeon machine, also using \texttt{AbsoluteTiming[]}. \textit{Internal} measured the time spent in the QuESTlink \texttt{C++} backend, and so excludes any Mathematica and WSTP overheads.
    Each datum is the average of $50$ tests with re-randomised gate parameters.
    }
    \label{fig:benchmark_gpu_dif_envs}
\end{figure}

To confirm this, we performed additional GPU benchmarking with a fixed circuit of 100 random rotations, and varying number of qubits. We study the overheads of local and remote emulation, using both \texttt{CreateLocalQuESTEnv[]} and \texttt{CreateRemoteQuESTEnv[]}.
The results are presented in Figure~\ref{fig:benchmark_gpu_dif_envs}, and reveal a negligible overhead on all platforms.

In general, use of a \textit{remote} QuEST environment will add a constant overhead to \QuESTlink's performance, due to network latency. Though in principle the cost of communicating a circuit from Mathematica to a remote backend scales linearly with the emulated circuit depth, this cost should be overshadowed by the exponentially growing cost of quantum emulation, and other network overheads. Indeed, this is exhibited in Figure~\ref{fig:benchmark_gpu_dif_envs}. The size of an encoded circuit is upper bounded by $8 \Sigma$\,bytes, where $\Sigma$ is the total number of gates, control qubits, target qubits and gate parameters (e.g. angles of rotation).
A circuit would need to contain approximately 30 million single-qubit gates to saturate a $1\,$GB bandwidth network.


\section{Future work}

\QuESTlink is currently under active development, with a growing list of planned work and new features. This list includes:

\begin{itemize}[leftmargin=1em]
\setlength\itemsep{0em}
    \item Integration with a GPU-accelerated linear algebra library, to perform fast numerical routines directly on the simulated quantum state. For example, diagonalisation of a density matrix enables rapid calculation of the Trace distance, which currently requires expensive communication with the Mathematica kernel.
    \item Support for emulation on remote, \textit{distributed} hardware.
    \item A QASM~\cite{openqasm} parser and generator, to and from \QuESTlink's circuit specification language.
\end{itemize}

We caution that \QuESTlink is still in an early form and likely to change, both in interface and architecture.


\section{Conclusion}

This manuscript introduced \QuESTlink, a Mathematica package for emulating quantum circuits, state-vectors and density matrices. \QuESTlink offers both high-level symbolic manipulation of quantum circuits, and rapid simulation using possibly remote hardware, such as multicore and GPU-accelerated supercomputers. We presented a broad technical overview of \QuESTlink, including its protocol for stand-alone installation-free deployment.
We then demonstrated the concision and flexibility possible of \QuESTlink, by stepping through several examples of otherwise sophisticated simulations.
These examples should enable an interested reader to begin using \QuESTlink immediately.
Lastly, we performed some simple benchmarks of \QuESTlink to estimate the overhead over core \QuEST, across its parallelisation modes. \QuESTlink is open-source, and accessible at \href{https://questlink.qtechtheory.org}{questlink.qtechtheory.org}, or on Github~\cite{questlink_github}

\section{Acknowledgements}

The authors sincerely thank Balint Koczor for help porting \QuESTlink to Windows, and performing extensive testing of \QuESTlink's analytic capabilities. We also thank
Suguru Endo and Balint Koczor for helping shape the functionality and direction of \QuESTlink and bug-spotting. Thanks also goes to the early adopters of \QuESTlink in the QTechTheory group, for their part in moulding a perfectly cromulent emulator.
TJ additionally thanks Quantum Motion Technologies Ltd for financial support in extending \QuESTlink, and Sinan Shi for his noble spirit and expertise in rubber duck debugging.
SCB acknowledges EPSRC grants EP/M013243/1 and EP/T001062/1, the IARPA funded project LogiQ, and the EU Flagship project AQTION. The authors would like
to acknowledge the use of the University of Oxford Advanced Research Computing facility (ARC, please see http://dx.doi.org/10.5281/zenodo.22558) in enabling the related QuEST work.

\bibliographystyle{unsrt}
\bibliography{bibliography}

\end{document}